\documentclass[a4paper,twoside,openany,11pt]{memoir} 

\def\fullheadfoot{0} 
\usepackage[english]{babel}
\usepackage{microtype,xspace}
\usepackage[utf8]{inputenc}	

\usepackage{amsfonts,amsmath,amssymb,bm,mathrsfs,mathtools,dsfont} 	
\allowdisplaybreaks 	
\usepackage[table]{xcolor}
\usepackage{hyperref}
\usepackage{slashed}
\usepackage{multicol}

\usepackage{siunitx}
\usepackage{geometry}
\usepackage[sort&compress,numbers,merge]{natbib}
\usepackage{chapterbib}
\addto\captionsenglish{\renewcommand*{\bibname}{References}}
\makeatletter
\renewcommand{\@memb@bchap}{ 
\bibmark \prebibhook
}
\makeatother

\usepackage{graphicx}
\graphicspath{{./Figures/}}
\usepackage{caption,subcaption}
\usepackage{multirow}

\usepackage{tabularx}
\newcolumntype{Y}{>{\centering\arraybackslash}X}

\usepackage[shortlabels]{enumitem}
\setlist{itemsep=.1em,topsep=.5em}
\SetEnumerateShortLabel{i}{\textit{\roman*}}

\definecolor{red}{rgb}{0.6,.0706,.1373}
\definecolor{blue}{rgb}{0,0.396,0.741}
\definecolor{green}{rgb}{0.25,0.6,0.2}
\definecolor{teal}{rgb}{0.11,0.6,0.6}
\definecolor{orange}{rgb}{.8, .4806, 0.173}
\definecolor{yellow}{rgb}{.8, .7, 0.05}
\colorlet{blueref}{blue!80!black}
\colorlet{bluelink}{blue!90!black}
\hypersetup{
	colorlinks, 
	bookmarksopen, 
	bookmarksnumbered,
	citecolor=blueref, 		
	linkcolor=bluelink,	
	urlcolor=bluelink,			
	pdftitle={A Partially Fixed Background Field Gauge},    
	pdfauthor={Anders Eller Thomsen},    	
}

\addto\captionsenglish{}

\renewcommand{\contentsname}{Contents}
\renewcommand{\printtoctitle}[1]{}

\settocdepth{subsection}
\newcommand{\toc}{ {
	\hypersetup{linkcolor = black} 
	\vspace*{-.06\textheight}	
	\tableofcontents*
	\thispagestyle{standardstyle} 
} }

\makeatletter
\newcommand*\ifthispageodd{%
  \checkoddpage
  \ifoddpage
    \expandafter\@firstoftwo
  \else
    \expandafter\@secondoftwo
  \fi
}
\makeatother

\usepackage[noindentafter]{titlesec} 
\counterwithout{section}{chapter} 
\counterwithout{figure}{chapter} 
\counterwithout{table}{chapter} 
\numberwithin{equation}{section} 
\setsecnumdepth{subsubsection}

\DeclareMathVersion{sans}
\SetSymbolFont{operators}{sans}{OT1}{cmbr}{m}{n}
\SetSymbolFont{letters}{sans}{OML}{cmbrm}{m}{it}
\SetSymbolFont{symbols}{sans}{OMS}{cmbrs}{m}{n}
\SetMathAlphabet{\mathit}{sans}{OT1}{cmbr}{m}{sl}
\SetMathAlphabet{\mathbf}{sans}{OT1}{cmbr}{bx}{n}
\SetMathAlphabet{\mathtt}{sans}{OT1}{cmtl}{m}{n}
\SetSymbolFont{largesymbols}{sans}{OMX}{iwona}{m}{n}

\DeclareMathVersion{boldsans}
\SetSymbolFont{operators}{boldsans}{OT1}{cmbr}{b}{n}
\SetSymbolFont{letters}{boldsans}{OML}{cmbrm}{b}{it}
\SetSymbolFont{symbols}{boldsans}{OMS}{cmbrs}{b}{n}
\SetMathAlphabet{\mathit}{boldsans}{OT1}{cmbr}{b}{sl}
\SetMathAlphabet{\mathbf}{boldsans}{OT1}{cmbr}{bx}{n}
\SetMathAlphabet{\mathtt}{boldsans}{OT1}{cmtl}{b}{n}
\SetSymbolFont{largesymbols}{boldsans}{OMX}{iwona}{bx}{n}

\titleformat{\section}{\centering \Large \bfseries \sffamily \mathversion{boldsans} \color{blue!80!black} }{\thesection}{15pt}{}{}
\titlespacing{\section}{0pt}{15pt}{5pt}
\titleformat{\subsection}{\large \sffamily \mathversion{sans} \color{blue!70!black} }{\thesubsection}{10pt}{}{}
\titlespacing{\subsection}{0pt}{10pt}{5pt}
\titleformat{\subsubsection}{\normalsize \sffamily \itshape \mathversion{sans} \color{blue!70!black} }{\thesubsubsection}{10pt}{}{}
\titlespacing{\subsubsection}{0pt}{10pt}{0pt}

\newcommand{\sectionlike}[1]{\phantomsection \addcontentsline{toc}{section}{#1} \setcounter{subsection}{0} \sectionmark{#1}
		\begin{center}
		\needspace{8\baselineskip}
		\Large \bfseries \sffamily \mathversion{boldsans} \color{blue!80!black} #1  
		\end{center}
	\vspace{-5pt} 
}

\ifnum\fullheadfoot=1
	\makepagestyle{standardstyle}
	\makeoddhead{standardstyle}{\sffamily \mathversion{subsectionmath} \rightmark}{}{\sffamily \bfseries{\thepage}}
	\makeevenhead{standardstyle}{\sffamily \bfseries{\thepage}}{}{\sffamily \mathversion{subsectionmath} \leftmark}
	\makeheadrule{standardstyle}{\textwidth}{\normalrulethickness}
	\makepsmarks{standardstyle}{
		\nouppercaseheads
		\createmark{section}{both}{shownumber}{\ }{\hspace{1.2em}  }
		\createmark{subsection}{right}{shownumber}{}{\hspace{1.2em} }
	}
	
	\copypagestyle{appstyle}{standardstyle}
	\makeevenhead{appstyle}{\sffamily \bfseries{\thepage}}{}{ \sffamily \mathversion{subsectionmath} \leftmark}
	\makepsmarks{appstyle}{
		\nouppercaseheads
		\createmark{section}{both}{nonumber}{\ }{\ \ }
		\createmark{subsection}{right}{shownumber}{}{\ \ }
	}
\else 
	\makepagestyle{standardstyle}
	\makeoddfoot{standardstyle}{}{\sffamily -- {\bfseries\thepage} --}{} 
	\makeevenfoot{standardstyle}{}{\sffamily -- {\bfseries\thepage} --}{}
	\copypagestyle{appstyle}{standardstyle}
\fi

\createplainmark{toc}{both}{\contentsname}
\createplainmark{bib}{both}{\bibname}

\makeatletter
\let\MyIntOrig\int
\def\MyIntSpace{\hspace{-.35em}} 
\def\int{\MyInt}
\def\MyInt{\MyIntOrig\MyIntSkipMaybe}
\def\MyIntSkipMaybe{
	\@ifnextchar_{\MyIntSkipScript}{%
		\@ifnextchar^{\MyIntSkipScript}{%
			\@ifnextchar\limits{\MyIntSkipTok}{%
				\@ifnextchar\nolimits{\MyIntSkipTok}{%
					\MyIntSpace}}}}%
}
\def\MyIntSkipScript#1#2{#1{#2}\MyIntSkipMaybe}
\def\MyIntSkipTok#1{#1\MyIntSkipMaybe}

\newcommand{\pushright}[1]{\ifmeasuring@#1\else\omit\hfill$\displaystyle#1$\fi\ignorespaces}
\makeatother

\newcommand{\brakets}[1]{\big\langle #1 \big\rangle}

\newcommand{\tr}{\mathop{\mathrm{Tr}} }
\newcommand{\eminus}{\vcenter{\hbox{\scalebox{0.6}[1]{$ - $}}}}	

\newcommand{\commutator}[2]{\big[#1, \, #2 \big]}

\newcommand{\hc}{\; + \; \mathrm{H.c.} \;}
\newcommand{\andeq}{\quad \mathrm{and} \quad}

\newcommand{\dd}{\mathop{}\!\mathrm{d}}
\newcommand{\ud}[2]{\phantom{}^{#1}\phantom{}_{#2}}
\newcommand{\du}[2]{\phantom{}_{#1}\phantom{}^{#2}} 
\newcommand{\diag}{\mathop{\mathrm{diag}}}
\newcommand{\transpose}{^{\intercal}}
\newcommand{\rep}[1]{\mathbf{#1}}
\newcommand{\repbar}[1]{\overline{\mathbf{#1}}}

\newcommand{\sscript}[1]{{\scriptscriptstyle \mathrm{#1}}}

\makeatletter
\newcommand{\vast}{\bBigg@{3}}
\makeatother

\renewcommand{\L}{\mathcal{L}}

\newcommand{\SU}{\mathrm{SU}}

\newcommand{\EFT}{\sscript{EFT}}

\definecolor{change}{rgb}{1,0.15,0.07}

\begin{document}

\thispagestyle{empty}
\renewcommand*{\thefootnote}{\fnsymbol{footnote}}
\suppressfloats	
\begin{center}
	{\sffamily \bfseries \fontsize{20}{22}\selectfont \mathversion{boldsans}
	A Partially Fixed Background Field Gauge
	\\[-.5em]
	\textcolor{blue!80!black}{\rule{.9\textwidth}{2pt}}\\
	\vspace{.05\textheight}}
	{\sffamily \mathversion{sans} \Large 
	  	Anders Eller Thomsen\footnote{anders.thomsen@unibe.ch}
	}\\[1.25em]
	{ \small \sffamily \mathversion{sans} 
	Albert Einstein Center for Fundamental Physics, Institute for Theoretical Physics,\\ University of Bern, CH-3012 Bern, Switzerland
	}
	\\[.005\textheight]{\itshape \sffamily \today}
	\\[.03\textheight]
\end{center}
\setcounter{footnote}{0}
\renewcommand*{\thefootnote}{\arabic{footnote}}%

\begin{abstract}\vspace{+.01\textheight}	
	We examine the incorporation of gauge symmetries in the modern effective field theory (EFT) matching paradigm with a particular focus on spontaneously broken symmetries. 
	The presence of gauge symmetries entails the introduction of gauge-fixing terms in matching calculations, which may prevent (partial) cancellation between loops from the underlying theory and those of the EFT, thereby preventing the establishment of a hard-region matching formula. 
	While this is not an issue when using the background field (BF) gauge for unbroken gauge theories, we find ourselves unable to demonstrate the cancellation with the ordinary BF gauge in spontaneously broken gauge theories when the massive vectors are heavy and decouple from the theory. As a convenient alternative, we construct a partially fixed BF gauge that, in addition to being simpler than the ordinary BF gauge, allows us to prove a hard-region matching formula. 
\end{abstract}

\newpage
\section*{Table of Contents}
\toc

\newpage 
\section{Introduction}

Gauge theories are the backbone of quantum field theory and an essential ingredient in the Standard Model (SM), although often the gauge symmetry is hidden rather than explicit. This spontaneous symmetry breaking (SSB) in gauge theories\footnote{Strictly speaking gauge symmetries are never broken, as they are mere redundancies of our mathematical description. We use the common nomenclature of spontaneous symmetry breaking to refer to the development of a Higgs phase in which (some of) the vector bosons become massive and only a remnant of the original gauge symmetry is manifest.} is a core mechanism of the SM and a frequent scenario in a myriad of models beyond the Standard Model (BSM), where new, postulated gauge symmetries are broken down to the SM gauge group or new fields alter the specifics of electroweak symmetry breaking. From the perspective of perturbative calculations, many of the difficulties of working with gauge theories stem from the necessity of gauge fixing the action, introducing both new interactions, ghost fields, and in many cases a proliferation of non-invariant counterterms. Furthermore, unphysical intermediate results may depend on the chosen gauge. These complications are further exacerbated in the event that the gauge symmetry is spontaneously broken. 

A common approach to studying the low-energy phenomenology of new BSM models is to use effective field theories (EFTs)~\cite{Dawson:2022ewj}, often in the form of the Standard Model Effective Field Theory (SMEFT)~\cite{Buchmuller:1985jz,Grzadkowski:2010es}. The idea is to match a BSM model to a suitable EFT, that is, to determine the action of the EFT such that it accurately reproduces the low-energy processes of the full model.
The fool-proof way to perform such a matching calculation is to compute a set of $ S $-matrix elements in both the EFT and the UV theory before equating them to fix the coefficients of the EFT operators. This approach works regardless of the gauge-fixing of the two theories and avoids all related subtleties. Unfortunately, it also entails computing physical amplitudes in both theories, which is very arduous, and requires a parametrization of the EFT Lagrangian with \emph{all} emerging operators ahead of time.

An alternative approach to EFT matching~\cite{Fuentes-Martin:2016uol,Zhang:2016pja,Fuentes-Martin:2023ljp} (see also~\cite{Manohar:1997qy}) based on the method of regions~\cite{Beneke:1997zp,Jantzen:2011nz} has become popular. It is based on an argument that the loop diagrams in the EFT are in one-to-one correspondence with the soft region of the loops in the UV theory, from which it follows that the EFT action can be identified with the hard region of the UV loops. This observation allows the enterprising EFT matcher to evaluate the EFT action directly, without computing loop diagrams in the EFT, and instead focus on the simpler hard-region UV loops. It should come as no surprise that this is the backbone of all (semi-)automated tools for one-loop EFT matching~\cite{Carmona:2021xtq,Fuentes-Martin:2022jrf,DasBakshi:2018vni,Cohen:2020qvb,Fuentes-Martin:2020udw}, incidentally, none of which have been used to match models with spontaneous symmetry breaking. 

The hard-region matching method stands and falls with the one-to-one correspondence of loop diagrams in the EFT and UV. 
This central claim of the method has been proven~\cite{Fuentes-Martin:2016uol,Zhang:2016pja,Fuentes-Martin:2023ljp} using the formalism of functional matching~\cite{Chan:1985ny,Gaillard:1985uh,Cheyette:1987qz,Henning:2014wua,Drozd:2015rsp,Boggia:2016asg,Ellis:2016enq,Kramer:2019fwz,Cohen:2020fcu,Dittmaier:2021fls} and, in turn, the hard-region matching method enables functional matching techniques. It is commonly held that background field (BF) gauges~\cite{DeWitt:1967ub,Abbott:1980hw,Abbott:1981ke} are suitable gauge fixings in functional techniques~\cite{Dittmaier:1995cr,Henning:2014wua} because they allow for the construction of a manifestly gauge-invariant effective action.\footnote{An early formulation~\cite{Cheyette:1987qz} relied instead on a background-gauge-invariant Landau gauge-fixing condition.} 
The BF gauges are, thus, used to directly extract a gauge-invariant EFT action, the goal of the matching exercise.
However, to our knowledge, no proof has been established of the required cancellation of one-loop EFT and UV diagrams that accounts for the intricacies of gauge-fixing. 
Even more so, in situations where SSB leads to a smaller gauge symmetry in the EFT compared to the UV theory, the one-to-one correspondence can seemingly not be directly established in the ordinary (linearly realized) BF gauge.\footnote{Our lack of success in proving the correspondence in the BF gauge is, on the other hand, not a proof that it cannot be made to work.}
An unrelated issue with using the BF gauge for matching in such SBB matching scenarios is the unmanageable mess that is the resulting gauge-fixing action. In an ideal world, such obstacles would be circumvented. 

In light of these challenges, this paper seeks to clarify gauge-fixing for matching calculations based on the method of regions. Our exploration allows us to demonstrate the soft-region cancellation of UV and EFT loops in BF gauges in models with unbroken gauge groups. This firmly establishes the validity of the approach that is widely used in the community. For the case of spontaneously broken gauge symmetries with decoupling massive vectors, we propose the use of a partially fixed gauge~\cite{Weinberg:1980wa,Ferrari:2013aza} for the heavy vectors in conjunction with a BF gauge for the remaining unbroken group. This factorizing combination leads to simpler and more manageable gauge-fixing terms in the action, while also allowing us to establish the much-vaunted correspondence between UV and EFT loops. By contrast, we will highlight the difficulties of establishing such a correspondence using the ordinary BF gauges for the full gauge group. 

The paper is organized as follows: Section~\ref{sec:BFM} reviews the background field method (BFM) for gauge fixing and how it can be applied to construct a gauge-invariant effective action. This subject may be familiar to some readers, but we emphasize a few subtleties that are often omitted in the literature. The discussion is extended to SSB in Section~\ref{sec:SSB}, which lays out the partial gauge-fixing framework and shows how it may be combined with the BF gauge for the unbroken remnant group. Finally, Section~\ref{sec:matching} proves the UV--EFT loop correspondence---so crucial to matching computations---for the case of BF gauge-fixing in models with unbroken gauge symmetries and for the case of the partially fixed BF gauge in models with SSB. We also discuss the challenges to establishing the correspondence in a regular BF gauge for the broken groups. We have also included several appendices to elaborate on various technical details of our discussion. 
The paper pursues a generic formulation of the problem such that our results may apply to all circumstances. This is particularly useful with an eye toward applications in the multitude of BSM models with all their unique gauge groups and breaking patterns. The unfortunate consequence of this choice is an admittedly dense notation, for which the reader will have to excuse us. We hope it is worth the trade-off.

\section{Gauge Fixing with the Background Field Method} \label{sec:BFM}
We begin our discussion by reviewing generating functionals for gauge theories and the role of gauge-fixing. This will allow us to establish the BF gauges and give a functional prescription for the gauge-invariant effective action up to one-loop order. This section will also introduce much of the notation used later on.

\subsection{Gauge theories and notation}
We seek to keep our discussion as general as possible, so our starting point will be a theory described by an action $ S[\eta] $, which is invariant under a gauge group $ G $.  The fields are collectively denoted $ \eta^I $, where $ I, J, \ldots $ run over all internal degrees of freedom. Obviously, the gauge fields $ A^A_\mu $ play a special role, so we write $ \eta^I = (A^A_\mu,\, \eta^p) $ with $ \eta^p $ denoting the matter fields. The indices $ A, B, \ldots $ are coordinates for the Lie algebra $ \mathfrak{g} $ associated with $ G $, and $ p, q,\ldots $ are collective indices for all remaining fields. 

The infinitesimal gauge transformation of the fields with gauge parameter $ \alpha = \alpha^A T_A \in \mathfrak{g} $ reads
	\begin{equation}
	\begin{aligned}
	\delta_\alpha A^A_\mu &\equiv D_\mu \alpha^A \equiv \partial_\mu \alpha^A + f\ud{A}{BC} A_\mu^B \alpha^C, \\
	\delta_\alpha \eta^p &= i \alpha^A T^p_{\!Aq} \eta^q.
	\end{aligned}
	\end{equation}
$ D_\mu $ denotes the ordinary covariant derivative, $ f\ud{A}{BC} $ is the structure constant of $ \mathfrak{g} $, and $ T^p_{Aq} $ a generator corresponding to the (reducible) representation of the matter fields.
We write the transformation compactly~\cite{Rebhan:1984bg} as 
	\begin{equation}
	\delta_\alpha \eta^I = D\ud{I}{C}[\eta] \alpha^C = \big(\delta\ud{A}{C} \partial_\mu + f\ud{A}{BC} A^B_\mu,\, i\, T^p_{\!Cq} \eta^q \big) \alpha^C,
	\end{equation}
where $ D\ud{I}{A}[\eta] $ is a differential operator, which is a linear function of $ \eta^I $, not to be confused with the covariant derivative $ D_\mu = \partial_\mu -i A_\mu $. Thus, $ D\ud{I}{A,J} = \delta D\ud{I}{A}[\eta]/ \delta \eta^J $ is independent of the field. 
The gauge action on the fields implied by the infinitesimal gauge transformation is compatible with the group structure given that\footnote{The condition implies that $ \delta_\alpha \delta_\beta \eta^I - \delta_\beta \delta_\alpha \eta^I= -i \delta_{[\alpha, \beta]} \eta^I  $.} 
	\begin{equation} \label{eq:variation_group_structure}
	D\ud{I}{A}[\eta](f\ud{A}{BC} \alpha^B \beta^C) = D\ud{I}{A,J} \beta^A D\ud{J}{B}[\eta] \alpha^B - D\ud{I}{A,J} \alpha^A D\ud{J}{B}[\eta] \beta^B, \qquad \alpha,\beta \in \mathfrak{g}.
	\end{equation}
Gauge invariance of the action, $ S[\eta + \delta_\alpha \eta] = S[\eta] $, implies that
	\begin{equation}\label{eq:eta_gauge_variation}
	S_{\!,I}[\eta] D\ud{I}{A}[\eta] = 0, \qquad 
	S_{\!,I}[\eta] \equiv \dfrac{\delta S}{\delta \eta^I}[\eta].
	\end{equation} 
The DeWitt summation convention employed here implies summation over the degrees of freedom in addition to integration over shared spacetime coordinates. We will also use DeWitt notation for functional derivatives elsewhere when it is unambiguous.

It is convenient to introduce a metric $ a^{\eminus 1}_{AB} $ on $ \mathfrak{g} $, incorporating the gauge couplings of the gauge group, along with an associated inner product. In terms of the basis vectors $ T_A $ of the Lie algebra, the inner product is\footnote{Gauge invariance dictates that $ f\ud{D}{CA} a^{\eminus 1}_{DB} + f\ud{D}{CB} a^{\eminus 1}_{AD} = 0 $.}   
	\begin{equation}
	\brakets{T_A,\, T_B} = a^{\eminus 1}_{AB}.
	\end{equation}
This metric has the cyclic property 
	\begin{equation}
	\brakets{T_A,\, \commutator{T_B}{T_C}} = i\,a^{\eminus 1}_{AD} f\ud{D}{BC} = \brakets{T_B,\, \commutator{T_C}{T_A}},
	\end{equation}
which is very useful. This follows from the coupling matrix being proportional to the identity matrix for any of the non-Abelian factors, along with the cyclic property of $ f\ud{A}{BC} $.
The Yang--Mills term of the gauge group is then 
	\begin{equation}
	\L_\sscript{YM} = - \dfrac{1}{4} \brakets{A_{\mu\nu},\, A^{\mu\nu}} = - \dfrac{1}{4} a_{AB}^{\eminus 1} A_{\mu \nu }^{A} A^{B\mu \nu}, 
	\end{equation}
where $ A_{\mu\nu} = i \commutator{D_\mu}{D_\nu} $ is the field strength tensor. 

The construction presented here does not make any assumptions about the gauge group. In general, $ G = \prod_n \! G_n $ can have any number of Abelian or simple factors. 
The diagonal elements of $ a^{\eminus 1}_{AB} $ take the value $ 1/g^2_n $  in terms of the coupling constants associated with the corresponding product groups. Meanwhile, gauge invariance ensures that off-diagonal elements vanish between non-Abelian factors. 
Lastly, off-diagonal elements between Abelian factors account for possible kinetic-mixing terms (see e.g., Ref.~\cite{Poole:2019kcm}). One might also consider a situation where the metric $ a_{AB}^{\eminus 1}[\eta] $ is taken to depend on the fields, such as in geometric theories. This generalization is briefly discussed in Appendix~\ref{app:curved_metric}.

\subsection{Gauge fixing and generating functionals}
It is well-known that gauge theories pose challenges to the naive application of perturbation theory; to avoid contributions from degenerate field configurations in the path integral and apply standard perturbative methods, the usual procedure is to fix the gauge.  
We consider a class of gauges, where the \emph{gauge-fixing condition} is parametrized by
	\begin{equation} \label{eq:BF_gauge_class}
	G^{A}[\eta,\, \Theta]= F\ud{A}{I}[\Theta] (\eta^I - \Theta^I),
	\end{equation}
for some auxiliary field $ \Theta $ and a differential \emph{gauge-fixing operator} $ F\ud{A}{I}[\Theta] $.
For instance, by identifying $ \Theta $ with the vacuum expectation value (VEV) of the fields, this class of gauges can reproduce the standard $ R_\xi $-gauges for models with or without SSB: $ G^A[\eta, \langle \eta \rangle ] = \partial^\mu A^A_\mu - i\, \xi \, a^{AB} \langle \eta^\ast_p \rangle T^p_{Bq} (\eta^q - \langle \eta^q \rangle) $.\footnote{For renormalization purposes, it is customary to include separate gauge parameters for each product group of~$ G $. This can be incorporated by replacing $ \xi \to \xi\ud{A}{B} $. We ignore this distraction, as the notation is plenty complex already.}
The gauge variation of the gauge-fixing condition reads 
	\begin{equation}
	\delta_\alpha G[\eta,\, \Theta] = F_{I}[\Theta] D\ud{I}{A}[\eta] \alpha^A. 
	\end{equation}

The partition function is gauge-fixed with $ G^A[\eta,\, \Theta] $ using the standard methods of Faddeev--Popov (summarized in Appendix~\ref{app:FP_gauge_fixing}): The gauge-fixing delta function $ \delta \big(G^A[\eta,\, \Theta] - \theta^A \big) $ is smoothed with a Gaussian weight over the $ \theta  $ parameter to obtain a renormalizable gauge-fixing term for the gauge fields. The fixing also introduces the functional determinant $ \mathrm{Det} \big( F\ud{A}{I}[\Theta] D\ud{I}{B}[\eta]  \big) $, which is interpreted as a Gaussian integral over the ghost and anti-ghost fields, collectively referred to as $ \boldsymbol{\omega}= (\omega^A,\, \overline{\omega}_A) $. 
Furthermore, we introduce a background field $ \overline{\eta}^I $ adding it to the quantum field $ \eta $ everywhere but in the source term of the fields. The result is the \emph{vacuum functional}\footnote{The shorthand $ \int_x \,= \int\, \dd^d x $ is employed throughout.} 
	\begin{equation}
	\mathcal{W} [J,\, \overline{\eta},\, \Theta] = 
	- i \log \! \int \mathcal{D} \eta \, \mathcal{D} \boldsymbol{\omega}\, \exp\! \left[i\left(\! S[\eta+ \overline{\eta}] + S^G_\mathrm{fix}[\eta + \overline{\eta},\, \boldsymbol{\omega},\, \Theta] + \int_x J_I \eta^I \right) \right],
	\end{equation}
where the gauge-fixing action for the gauge group $ G $ is given by
	\begin{equation}
	S^G_\mathrm{fix}[\eta,\, \boldsymbol{\omega},\, \Theta] = - \int_x \bigg( \dfrac{1}{2\xi} \big\langle F_I[\Theta] (\eta -\Theta)^I,\, F_J[\Theta] (\eta -\Theta)^J \big \rangle + \overline{\omega}_A F\ud{A}{I}[\Theta] D\ud{I}{B}[\eta]\, \omega^B
	\bigg).
	\end{equation}
$ \mathcal{W} $ is the generating functional of all connected Green's functions of the theory with field sources $ J_I $ (in the presence of $ \overline{\eta} $ and $ \Theta $). It can be used to determine the physical $ S $-matrix with standard methods, the result of which is independent of the background fields~\cite{Becchi:1999ir,Ferrari:2000yp}. 	

The \emph{quantum effective action}, the generating functional for one-particle irreducible (1PI) Green's functions, is obtained by the usual Legendre transformation of the vacuum functional:
	\begin{equation} \label{eq:conventional_eff_action}
	\Gamma[\hat{\eta},\,\overline{\eta} ,\, \Theta] = \mathcal{W}[J,\, \overline{\eta}, \, \Theta] - \int_x J_I \hat{\eta}^I, \qquad
	\hat{\eta}^I = \dfrac{\delta \mathcal{W}[J,\,\overline{\eta} ,\, \Theta]}{\delta J_I}, \qquad 
	-J_I = \dfrac{\delta \Gamma[\hat{\eta},\,\overline{\eta},\, \Theta]}{\delta \hat{\eta}^I},
	\end{equation}
trading the field source $ J $ for the expectation value of the field, $ \hat{\eta} $ (sometimes called the classical field), in the presence of the source.
This is typically the simplest object to work with for the purposes of RG calculations, as it contains all the information of the vacuum functional but with fewer Feynman diagrams. By shifting the integration variable from $ \eta \to \eta + \hat{\eta} $, we arrive at the path integral
	\begin{equation} \label{eq:gauge-fixed_BF_vacuum_functional}
	\Gamma[\hat{\eta},\, \overline{\eta} ,\, \Theta] = -i \log \!  
	\int \mathcal{D} \eta \, \mathcal{D} \boldsymbol{\omega}\, \exp\! \left[i\left(\! S[\eta+ \overline{\eta} + \hat{\eta}] + S^G_\mathrm{fix}[\eta + \overline{\eta}+ \hat{\eta},\, \boldsymbol{\omega},\, \Theta] + \int_x J_I \eta^I \right) \right].
	\end{equation}
In the vanilla formulation of the effective action, there are no background fields and the 1PI Green's functions are generated by the functional $ \Gamma[\hat{\eta},\, 0 ,\, \Theta=0] $. However, since the effective action~\eqref{eq:gauge-fixed_BF_vacuum_functional} depends on sources only in the combination $ \overline{\eta} + \hat{\eta} $, it is obvious that 
	\begin{equation} \label{eq:bfm_vs_conventional_eff_action}
	\Gamma[0,\, \overline{\eta},\, \Theta] = \Gamma[\hat{\eta} = \overline{\eta},\,0 ,\, \Theta].
	\end{equation}
This allows for evaluating the effective action through the use of the background field alone---what is known as the \emph{background field method} (BFM) for the effective action. It contains all the 1PI diagrams with external $ \overline{\eta} $ legs. 
The effective action is, however, not gauge invariant for generic gauges, and retains only the smaller BRST symmetry. This symmetry reduction is rather inconvenient for many kinds of calculations; for one, renormalization requires many new counterterms.

\subsection{The gauge-invariant effective action} \label{sec:gauge-invariant_eff_action}
The background field gauge is not so much a gauge in the conventional sense as it is a particular variation of the path integral cleverly constructed to produce a gauge-invariant effective action. A key ingredient is the choice of a gauge-fixing operator $ F\ud{A}{I}[\Theta] $ satisfying
	\begin{equation} \label{eq:gauge_op_transformation}
	\big( D\ud{J}{B}[\Theta] \alpha^B \big) F\ud{A}{I,J}[\Theta] X^I +  F\ud{A}{J}[\Theta] (D\ud{J}{B,I} \alpha^B X^I)= - f\ud{A}{BC} \alpha^B  F\ud{C}{I}[\Theta] X^I 
	\end{equation}
for any $ X^I $ and $ \alpha \in \mathfrak{g} $, that is, $ F\ud{A}{I}[\Theta] $ must transform covariantly w.r.t. a transformation of $ \Theta^I \to D\ud{I}{A}[\Theta] \alpha^A $ and $ X^I \to D\ud{I}{A,J} \alpha^A X^J $.\footnote{In contrast to $ \Theta $, $ X $ is taken to transform homogeneously. This reflects the response of $ \eta $ to a background field gauge transformation, which is introduced shortly.} This property of $ F $ is ensures the required invariance of the result. Next, we choose for our auxiliary field $ \Theta = \overline{\eta} $, to involve the background field in the gauge-fixing condition. We retain a generic condition for our discussion here to maintain generality; the reader may instead think of $ F\ud{A}{I}[\overline{\eta}] \eta^I = \overline{D}^\mu A_\mu^A  $ as a concrete and commonly used choice for the gauge-fixing operator ($ \overline{D}_\mu = \partial_\mu - i \overline{A}_\mu $ is a covariant derivative but with the background rather than the quantum gauge field). 

As mentioned, we proceed to fix the gauge of the quantum field using the background field of the BFM. The resulting class of gauges is called the \emph{background field gauges}~\cite{DeWitt:1967ub,Abbott:1980hw,Abbott:1981ke}. We, therefore, consider the vacuum functional 
	\begin{equation} \label{eq:BFM_W-tilde}
	\widetilde{\mathcal{W}}[J,\, \overline{\eta}] \equiv  \mathcal{W}[J,\, \overline{\eta},\, \Theta= \overline{\eta}]. 
	\end{equation}
The gauge-fixing term takes a particularly simple form in this case:
	\begin{equation} \label{eq:BFM_fixing_term}
	S^G_\mathrm{fix}[\eta + \overline{\eta},\, \boldsymbol{\omega},\, \overline{\eta}] = - \int_x \bigg( \dfrac{1}{2\xi} \big\langle F_I[\overline{\eta}] \eta^I,\, F_J[\overline{\eta}] \eta^J \big \rangle + \overline{\omega}_A F\ud{A}{I}[\overline{\eta}] D\ud{I}{B}[\eta+ \overline{\eta}]\, \omega^B
	\bigg).
	\end{equation}
One can then show that $ S^G_\mathrm{fix}[\eta + \overline{\eta},\, \boldsymbol{\omega},\, \overline{\eta}] $ and, therefore, $ \widetilde{\mathcal{W}}[J,\, \overline{\eta}] $ are invariant under the \emph{background gauge transformation} defined by
	\begin{equation}
	\overline{\delta}_\alpha \overline{\eta}^I = D\ud{I}{A}[\overline{\eta}] \alpha^A, \qquad \overline{\delta}_\alpha J_I = -J_J D\ud{J}{A,I} \alpha^A,
	\end{equation} 
which may be verified by also shifting the integration variable---the quantum field---by $ \overline{\delta}_\alpha \eta^I = D\ud{I}{A,J} \eta^J \alpha^A$,\footnote{
For illustration, this means that the background gauge transformation acts as $ \overline{\delta}_\alpha \overline{A}^A_\mu = \partial_\mu \alpha + f\ud{A}{BC} \overline{A}^B_\mu \alpha^C $ and $ \overline{\delta}_\alpha A^A_\mu = f\ud{A}{BC} A^B_\mu \alpha^C $ for background and quantum gauge fields, respectively.
} that is, the quantum fields are taken to transform homogeneously. This relies on the transformation properties of the gauge-fixing operator~\eqref{eq:gauge_op_transformation} and the group structure of the gauge variation~\eqref{eq:variation_group_structure}. 
Invariance the action $ S[\eta + \overline{\eta}] $ follows from ordinary gauge invariance of $ S[\eta] $ along with the observation that $ \overline{\delta}_\alpha (\eta + \overline{\eta})^I = D\ud{I}{A}[\eta + \overline{\eta}] \alpha^A $.

We can express the invariance of the vacuum functional under background gauge transformations with the identity
	\begin{equation} \label{eq:W-tilde_invariance}
	0 = \overline{\delta}_\alpha \widetilde{\mathcal{W}}[J,\, \overline{\eta}] =  \dfrac{\delta \widetilde{\mathcal{W}}[J,\, \overline{\eta}]}{\delta \overline{\eta}^I} D\ud{I}{A}[\overline{\eta}] \alpha^A - J_I \alpha^A D\ud{I}{A,J} \dfrac{\delta \widetilde{\mathcal{W}}[J,\, \overline{\eta}]}{\delta J_J}.
	\end{equation}
We proceed to construct the associated effective action
	\begin{equation}
	\widetilde{\Gamma}[\tilde{\eta},\, \overline{\eta}] = \widetilde{\mathcal{W}}[J,\, \overline{\eta}] - \int_x J_I \tilde{\eta}^I = \Gamma[\tilde{\eta},\, \overline{\eta},\, \overline{\eta}],
	\qquad \tilde{\eta}^I = \dfrac{\delta \widetilde{\mathcal{W}}[J,\, \overline{\eta}]}{\delta J_I},
	\qquad - J_I = \dfrac{\delta \widetilde{\Gamma}[\tilde{\eta}, \overline{\eta}]}{\delta \tilde{\eta}^I}.
	\end{equation}
Treating $ \tilde{\eta} $ and $ \overline{\eta} $ as independent variables, Eq.~\eqref{eq:W-tilde_invariance} results in  
	\begin{equation} \label{eq:gauge_invariance_Gamma-tilde}
	\begin{split}
	0 = \overline{\delta}_\alpha \widetilde{\Gamma}[\tilde{\eta}, \, \overline{\eta}]= \dfrac{\delta \widetilde{\Gamma}[\tilde{\eta},\, \overline{\eta}]}{ \delta \tilde{\eta}^I } D\ud{I}{A,J} \alpha^A \tilde{\eta}^J +   \dfrac{\delta \widetilde{\Gamma}[\tilde{\eta},\, \overline{\eta}]}{ \delta \overline{\eta}^I } D\ud{I}{A}[\overline{\eta}] \alpha^A.
	\end{split}
	\end{equation}
Not surprisingly, we conclude that also $ \widetilde{\Gamma} $ is (background) gauge invariant. Unfortunately, the condition $ \overline{\eta} = 0 $ is not background gauge invariant, so $ \widetilde{\Gamma}[\tilde{\eta}, 0] $ cannot be used to build invariant Green's functions of $ \tilde{\eta} $ to connect to physical amplitudes in the absence of $ \overline{\eta} $.

Inspired by the BFM~\eqref{eq:bfm_vs_conventional_eff_action}, we are lead to define our \emph{gauge-invariant effective action}
	\begin{equation} \label{eq:gauge-invariant_effective_action}
	\overline{\Gamma}[\overline{\eta}] \equiv \widetilde{\Gamma}[0,\, \overline{\eta}] = \Gamma[0,\, \overline{\eta}, \, \Theta= \overline{\eta}]
	\end{equation}
as a generating functional of 1PI Green's functions. It is evident that $ \overline{\Gamma}[\overline{\eta}] $ generates new vertices compared to the effective action of the conventional gauge $ \Gamma[\hat{\eta},\, 0,\, \Theta] $, namely those generated by taking $ \Theta $ derivatives before fixing the auxiliary field; however, $ \overline{\Gamma} $ still contains the physics of the theory. Eq.~\eqref{eq:gauge_invariance_Gamma-tilde} clearly implies the gauge invariance of the $ \overline{\Gamma} $:
	\begin{equation}
	0 = \overline{\delta}_\alpha \overline{\Gamma}[\overline{\eta}] = \overline{\Gamma}_{\!, I}[\overline{\eta}] D\ud{I}{A}[\overline{\eta}] \alpha^A. 
	\end{equation}
Thus, the gauge-invariant effective action satisfies Ward identities associated with the preserved background gauge symmetry. This has also been explicitly verified in the case of the SM and SMEFT~\cite{Denner:1994xt,Corbett:2020ymv}.

Due to the gauge invariance, the 2-point functions of the background gauge field produced by $ \overline{\Gamma} $ are singular (just as for the gauge-invariant classical action); hence, $ \overline{\Gamma} $ cannot be directly Legendre-transformed into a generating functional for connected Green's functions. A solution is to add an independent gauge-fixing term for the background fields:
	\begin{equation}
	\Gamma_\sscript{BF}[\overline{\eta}] = \overline{\Gamma}[\overline{\eta}] + S_\mathrm{bg.}^G[\overline{\eta}], \qquad 
	S_\mathrm{bg.}^G[\overline{\eta}] = \int_x \big\langle \mathcal{F}_I[\overline{\eta}] \overline{\eta}^I ,\, \mathcal{F}_J[\overline{\eta}] \overline{\eta}^J \big\rangle,
	\end{equation}    
where $ \mathcal{F}_I \in \mathfrak{g} $ is an arbitrary gauge-fixing operator that can be chosen independently of $ F_{I} $. By a Legendre transformation, we arrive at a generating functional
	\begin{equation} \label{eq:W_BF_definition}
	\mathcal{W}_\sscript{BF}[\overline{J}] = \Gamma_\sscript{BF}[\overline{\eta}] + \int_x \overline{J}_{I} \overline{\eta}^I, \qquad \overline{J}_{I} = -\dfrac{\delta \Gamma_\sscript{BF}[\overline{\eta} ]}{\delta \overline{\eta}^I},  
	\end{equation}
for the connected Green's functions in the BF gauges. We stress that off-shell $ \mathcal{W}_\sscript{BF} $ does not exactly reproduce $ \mathcal{W} $ of a conventional gauge---indeed, the off-shell vacuum functionals associated with conventional gauges are known to be dependent on the gauge choice. Nevertheless, it has been shown that the physical $ S $-matrix resulting from the Green's functions generated by $ \mathcal{W}_\sscript{BF}[\overline{J}] $ is identical to that obtained by using a conventional gauge, i.e., from $ \mathcal{W}[J,\, \Theta] $~\cite{Rebhan:1984bg,Abbott:1983zw,Hart:1983lbv,Becchi:1999ir,Ferrari:2000yp}.

\subsection{One-loop gauge-invariant effective action} \label{sec:one-loop_eff_action}
Most matching calculations in the BSM context are performed at no higher than one-loop order. To eventually discuss the matching of gauge theories and show a partial cancellation of EFT and UV loops, we will need a more explicit expression for the one-loop effective action. The functional form of the effective action~\cite{Jackiw:1974cv,Iliopoulos:1974ur} has proven very useful for this purpose and and will be used extensively in Section~\ref{sec:matching}. Let us introduce the notation 
	\begin{equation}
	S^{\mathrm{all}}[\eta, \, \boldsymbol{\omega},\, \Theta] = S[\eta] + S_\mathrm{fix}^{G}[\eta, \, \boldsymbol{\omega},\, \Theta]
	\end{equation} 
for compactness and denote the solution to the tree-level EOM
	\begin{equation}
	\dfrac{\delta S^{\mathrm{all}}}{\delta \eta^I}[\underline{\eta},\, 0,\, \Theta] = - J_I
	\end{equation}
by $ \underline{\eta} $ (conservation of the ghost number implies that $ \underline{\boldsymbol{\omega}} =0 $).

The one-loop effective action can be determined with a saddlepoint approximation of the path integral.
Shifting the integration variable $ \eta \to \eta + \underline{\eta} - \overline{\eta} - \hat{\eta} $, the one-loop effective action~\eqref{eq:gauge-fixed_BF_vacuum_functional} becomes  
	\begin{equation}
	\Gamma[\hat{\eta},\, \overline{\eta},\, \Theta] = S^{\mathrm{all}}[\underline{\eta},\, 0,\, \Theta] + \dfrac{i}{2} \mathrm{sTr} \log \mathcal{Q}_{\mathbf{IJ}}^{\mathrm{all}}[\underline{\eta},\, \Theta] + J_I (\underline{\eta} - \overline{\eta} - \hat{\eta})^I,
	\end{equation}
where the functional supertrace accounts for the mixed statistics of the quantum fields (see, e.g., Refs.~\cite{Zhang:2016pja,Cohen:2020fcu}). 
The supertrace effectively sums over all quantum degrees of freedom (DOFs), which in a diagrammatic language corresponds to the sum of all one-loop diagrams. The fluctuation operator is given by 
	\begin{equation} \label{eq:fluct_op}
	\mathcal{Q}^{\mathrm{all}}_{\mathbf{IJ}}[\underline{\eta},\, \Theta] = \dfrac{\delta^2 S^{\mathrm{all}}}{\delta \Omega^\mathbf{I} \delta \Omega^\mathbf{J}}[\underline{\eta},\, 0,\, \Theta], \qquad \Omega^\mathbf{I} = (\eta^I,\, \overline{\omega}_A,\, \omega^A),
	\end{equation}
where the indices run over all quantum DOFs in $ S^{\mathrm{all}} $, including the ghost fields, at all spacetime points. We have been forced to make a distinction between bold and italicized indices: there are only so many unique alphabets to use.
The quantum EOM for $ \hat{\eta} $, given in the last equality of Eq.~\eqref{eq:conventional_eff_action}, indicates that in the tree-level approximation $ \underline{\eta} = \overline{\eta} + \hat{\eta} $, from which it follows that
	\begin{equation}
	\Gamma[\hat{\eta},\, \overline{\eta},\, \Theta] = S^{\mathrm{all}}[\overline{\eta} + \hat{\eta},\, 0,\, \Theta] + \dfrac{i}{2} \mathrm{sTr} \log \mathcal{Q}_{\mathbf{IJ}}^{\mathrm{all}}[\overline{\eta} + \hat{\eta},\, \Theta].
	\end{equation}
Going to the BF gauges, the gauge-invariant effective action takes the particularly simple form 
	\begin{equation} \label{eq:one-loop_GI_effecitve_action}
	\overline{\Gamma}[\overline{\eta}] = S[\overline{\eta}] + \dfrac{i}{2} \mathrm{sTr} \log \mathcal{Q}_{\mathbf{IJ}}^{\mathrm{all}}[\overline{\eta},\, \overline{\eta}]
	\end{equation}
up to one-loop order, as $ S_\mathrm{fix}^{G}[\overline{\eta}, \, 0,\, \overline{\eta}] = 0 $. 

Casting the one-loop effective action as the supertrace of a functional log is a useful way to probe it from a theory perspective, although it is not immediately clear how it might be evaluated. In general, non-localities of the loop integrals embedded in the trace prevents the derivation of a closed-form formula. An expansion parameter is needed in the form of the number of external fields or powers of the couplings of the theory. For RG calculation, such as EFT matching or the calculations of $ \beta $-functions, one can focus on the hard region of the functional trace, which lends itself especially well to evaluation~\cite{Ellis:2016enq,Kramer:2019fwz,Cohen:2020fcu,Dittmaier:2021fls,Cohen:2020fcu,Fuentes-Martin:2020udw}.

\section{Spontaneous Symmetry Breaking}\label{sec:SSB}
Having reviewed gauge-fixing and the BF gauges for unbroken gauge theories, we are now ready to tackle the more complicated task of gauge-fixing spontaneously broken gauge symmetries. If one were to use an ordinary BF gauge, it is possible to once again construct an effective action that is invariant under the full gauge group albeit at the price of what is potentially very complicated gauge-fixing terms (cf. Appendix~\ref{app:broken_phase_BF-gauge}). Instead, we seek to construct an effective action, which is invariant only under transformation from the unbroken remnant group.

\subsection{Gauge theories in the broken phase} \label{sec:broken_gauge_theory}
We consider the case where among the matter fields there are scalars, collectively referred to as $ \varphi^{\prime a} \subset \eta^p $, that develop a VEV $ v^a = \langle \varphi' \rangle^a $. The result is a breaking of the gauge group to the remnant group $ H\subseteq G $, which is the largest subgroup that leaves the VEV invariant.
We let\footnote{The factor $ -i $ in $ f $ is due to using Hermitian generators for the scalar field following the conventions of Ref.~\cite{Peskin:1995ev}. Thus, $ T^a_{Ab} $ is purely imaginary for real representations.}  
	\begin{equation} \label{eq:decay_constant}
	\varphi^{\prime a} = v^a + \varphi^a \andeq f\du{A}{a} = -iT_{\!Ab}^{a} v^b
	\end{equation}
such that $ f\du{A}{a} $ is the matrix of decay constants in the gauge basis (not to be confused with the structure constants $ f\ud{A}{BC} $ with three indices). $ T^a_{Ab} $ denotes the appropriate representation of the generators of $ G $. Rather than restricting the formulation to the use of real scalar fields, complex fields can be accommodated in $ \varphi^{\prime a} $ by including both degrees of freedom. In an example where the theory contains a real and a complex scalar field---$ \varphi_1 $ and $ \varphi_2 $, respectively---we would write $ \varphi^a = (\varphi_1, \varphi_2, \varphi_2^\ast) $. In this construction, a metric $ h_{ab} $ is needed to construct a gauge singlet from the scalar field. Thus, for the example at hand $ \varphi^a h_{ab} \varphi^b = \varphi^2_1 + 2 \varphi_2^\ast \varphi_2 $. In general, we may always choose the $ h_{ab} $ as a matrix with 1's on the diagonal except for pairs of complex fields, which get two off-diagonal 1's instead. Another viable construction is to choose $ \varphi^{\prime a} $ purely real,\footnote{Complex scalar fields can be written in terms of real components, so this is always possible.} in which case we would simply identify $ h_{ab} =\delta_{ab} $. 

With the scalars in place, we observe that the scalar kinetic term gives rise to the familiar mass term for the vectors, which reads
	\begin{equation}
	\L_\mathrm{kin} = \tfrac{1}{2} D_\mu \varphi^{\prime}_a D_\mu \varphi^{\prime a} \supset \tfrac{1}{2} A_\mu^A (f h f\transpose)_{AB} A^{B\mu}, \qquad \varphi'_a = h_{ab} \varphi^{\prime b}.
	\end{equation}
Both here and elsewhere $ h_{ab} $ is used to lower scalar-type indices. 
The vectors are brought into the mass basis with the transformation 
	\begin{equation} \label{eq:L_transformation}
	A^{A}_\mu \longrightarrow  L\ud{A}{B} A_{\mu}^B = \binom{B^\alpha_\mu}{V^i_\mu}, \qquad T_{A} \longrightarrow T_{B} (L^{\eminus 1})\ud{B}{A} = \binom{t_{\alpha}}{x_{i}}, \qquad L = \hat{a}^{ 1/2} U^\dagger \, a^{\eminus 1/2}
	\end{equation}
such that $ B^\alpha_\mu $ is the collection of massless gauge fields associated with $ H $ and $ V^i_\mu $ the multiplet of massive vectors. Similarly, we may discriminate between the generators $ t_\alpha $ and $ x_i $ associated with the unbroken and broken directions in $ \mathfrak{g} $, respectively. The gauge index decomposes as $ A = \{ \alpha, \,i\} $. Hence, we may write the covariant derivative in the mass basis as
	\begin{equation} \label{eq:covD_heavy-light_split}
	D_\mu = \partial_\mu - i B_\mu^\alpha t_\alpha - i V_\mu^i x_i = d_\mu - i V_\mu^i x_i,
	\end{equation}
which introduces $ d_\mu $ as the covariant derivative w.r.t. the unbroken group, $ H $. 

The construction of $ L\ud{A}{B} $ in Eq.~\eqref{eq:L_transformation} is obtained by first rescaling the gauge couplings away from the Yang--Mills term,\footnote{The notation $ a^{\eminus 1/2} $ a mnemonic. It is simply some matrix satisfying $ a^{\eminus 1/2} a (a^{\eminus 1/2})\transpose = \mathds{1} $,  which in the typical case of a diagonal coupling matrix $ a $ can be chosen as the matrix with the inverse square root of those same couplings on the diagonal. In the event of kinetic mixing, it is often convenient to use a triangular matrix to remove these off-diagonal terms. The same applies for the notation $ \hat{a}^{1/2} $, which satisfies $ \hat{a}^{1/2} \diag(\mathds{1},\, \kappa) (\hat{a}^{1/2})\transpose = \hat{a} $.} then doing a unitary transformation $ U $ to canonize the resulting mass term, before rescaling the massless fields with their couplings through $ \hat{a}^{1/2} $, such that the gauge couplings associated with $ H $ appear in their kinetic term.  
Choosing an orthogonal $ U $ would be sufficient to diagonalize the mass term of the massive vectors; however, since the heavy vectors often include complex representations under $ H $, we allow for embedding the complex fields in $ V_\mu $ in a manner similar to how the complex scalar representations are embedded in $ \varphi' $. To accommodate this situation, $ U $ is generically a unitary matrix satisfying 
	\begin{equation}
	(U\transpose U)_{\alpha\beta} = \delta_{\alpha \beta}, \qquad (U\transpose U)_{ij} = \kappa_{ij}, \qquad (U\transpose U)_{\alpha j} = (U\transpose U)_{i\beta} = 0,
	\end{equation}
where $ \kappa_{ij}= \kappa^{ij} $ is a unitary metric on the algebra of the broken vectors, which consists of 1's on the diagonals corresponding to the real fields and pairwise off-diagonal 1's for the complex pairs. $ \hat{a} $ is a block diagonal matrix satisfying
		\begin{equation}
		\hat{a}^{ij} = \kappa^{ij}, \qquad 
		\hat{a}^{i\alpha}= \hat{a}^{\alpha i} = 0, \qquad 
		\hat{a}^{\alpha \beta} = (L a L\transpose)^{\alpha \beta}.
		\end{equation}
It takes the value $ \kappa $ on the space of the broken vectors, while it is identified with the coupling matrix of the gauge group $ H $ on $ \mathfrak{h} \subseteq \mathfrak{g}$, the Lie algebra of $ H $.

With the transformation sketched here, the masses $ M_i $ (the mass index is not counted in the summation convention) of the massive gauge fields appear in the canonized mass matrix
	\begin{equation} \label{eq:heavy_V_mass_matrix}
	M_{i}^{2} \kappa_{ij} = [L^{\eminus 1 \intercal} f h  f\transpose L^{\eminus 1}]_{ij} = [U\transpose a^{1/2} f h f\transpose a^{1/2} U]_{ij}.
	\end{equation}
The metric on the new generators is found to be 
	\begin{equation}
	\brakets{t_{\alpha}, t_{\beta}} = L\ud{\eminus 1\, A}{\alpha} L\ud{\eminus 1\, B}{\beta} \brakets{T_{A},\, T_{B}} = \hat{a}^{\eminus 1}_{\alpha \beta},
	\end{equation}
which defines the inner product on $ \mathfrak{h} $. 
Similarly, we have\footnote{The ``broken'' generators, $ x^i $, are to be understood as some gauge coupling times a generator.}
	\begin{equation}
	\brakets{x_{i}, t_{\alpha}} = 0, \andeq \brakets{x_{i},\, x_j} = \kappa_{ij}.
	\end{equation}
We will at times use $ \kappa_{ij} $ and its inverse $ \kappa^{ij} = \kappa_{ij} $ to raise and lower the indices in the broken part of the algebra (similarly to the use of $ h_{ab} $). On the other hand, $ \hat{a}^{\alpha \beta} $ contains the couplings of the remnant group and is kept explicit throughout our discussion. 

In a slight abuse of notation we retain the use of $ f $, but with Greek indices $ \alpha, \beta, \ldots $ and lowercase Latin indices $ i,j,\ldots $ for the structure constants transformed with $ L $ to the mass basis:
	\begin{equation}
	f\ud{\alpha(i)}{\beta(j)\,\gamma(k)} = L\ud{\alpha(i)}{A} f\ud{A}{BC} (L^{\eminus 1})\ud{B}{\beta(j)} (L^{\eminus 1})\ud{C}{\gamma(k)}.
	\end{equation}
Thus, $ f\ud{\alpha}{\beta\gamma} $ is the structure constant of $ H $. Due to group closure, the symbols with two Greek and one Latin index vanish, e.g., $ f\ud{\alpha}{\beta i} =0 $. The rest are expected to be non-zero in most situations. A similar notation is employed for the decay constant 
	\begin{equation}
	f\du{\alpha(i)}{a} = (L^{\eminus 1})\ud{A}{\alpha(i)} f\du{A}{a},
	\end{equation}
where $ f\du{\alpha}{A} =0 $. Additional details regarding the construction of the Lagrangian in the broken phase are provided in Appendix~\ref{app:broken_phase} along with mass basis expressions for the scalar and gauge kinetic terms, while Appendix~\ref{app:toy_model} illustrates how to work out the various quantities in a toy model.

\subsection{Partial gauge fixing}
In the spontaneously broken gauge theory, it is still necessary to include gauge-fixing to remove singularities from the kinetic terms of the massive vectors. With the development of a scalar VEV, we loose explicit invariance of the action under the full gauge group. 
Here we wish to fix the gauge in the direction of $ G/H $ in such a way as to maintain explicit invariance w.r.t. $ H $ gauge transformations. This partial gauge fixing was proposed by Weinberg~\cite{Weinberg:1980wa} for the purposes of integrating out the heavy vectors and constructing the low-energy EFT. Unfortunately, his construction relied on fixing the gauge with a delta function without the Gaussian weight integration used in standard renormalizable gauges, such as the $ R_\xi $-gauges. Ferrari~\cite{Ferrari:2013aza} resolved the issue and described in some detail how to reintroduce the Gaussian weight without losing invariance under $ H $.  

We write the gauge-fixing condition for $ G $ as $ L\ud{A}{B} G^{B}[\eta] = (G^\alpha[\eta],\, G^{i}[\eta] ) $, where $ G^\alpha $ fixes the gauge for the gauge fields of $ H $ and $ G^i $ fixes the massive vectors associated to $ G/H $. The idea is to choose $ G^\alpha $ to be independent of all the massive vectors, which leads to similarly independent ghost terms, since $ H $ closes under group multiplication. The heavy vectors $ V^i_\mu $ are in a representation of $ H $, which lets us choose a $ G^i $ that transforms covariantly under $ H $:
	\begin{equation} \label{eq:heavy_gf_H-invariance}
	\delta^H_\alpha G^i[\eta] =  G\ud{i}{\!,I}[\eta] D\ud{I}{\alpha}[\eta] \alpha^{\alpha} = - f\ud{i}{\alpha j} G^{j}[\eta] \alpha^{\alpha}.
	\end{equation}
This is easily achieved in practice by using derivatives that are covariant w.r.t. the remnant group and we will provide a concrete choice in Section~\ref{sec:part_fixed_gf_conditions}. 

The covariant transformation of $ G^i $ ensures that the integral of the gauge-fixing delta function over the gauge group~\eqref{eq:int_delta_gf} evaluates to
	\begin{equation}
	\begin{split}
	\int_{G} \dd \mu[g]\, \delta\big(G^{A}[\eta_g]\big) 
	&= \mathrm{Det}^{\eminus 1}\! \begin{pmatrix}
		G\ud{\alpha}{\!,I}[ \eta_{g_0}] D\ud{I}{\beta}[ \eta_{g_0}] &
		G\ud{\alpha}{\!,I}[ \eta_{g_0}] D\ud{I}{j}[ \eta_{g_0}] \\
		 - f\ud{i}{\beta k} G^{k}[\eta_{g_0}] &  
		G\ud{i}{\!,I}[ \eta_{g_0}] D\ud{I}{j}[ \eta_{g_0}]
		\end{pmatrix}\\
	&= \mathrm{Det}^{\eminus 1} \big( G\ud{\alpha}{\!,I}[ \eta_{g_0}] D\ud{I}{\beta}[ \eta_{g_0}] \big) \mathrm{Det}^{\eminus 1} \big( G\ud{i}{\!,I}[ \eta_{g_0}] D\ud{I}{j}[ \eta_{g_0}] \big),
	\end{split}
	\end{equation} 
as $ G^{j}[\eta_{g_0}] =0$. Here $ \eta_g $ denotes the action of $ g\in G $ on the fields $ \eta $, while $ g_0 $ denotes the solution to $ G^A[\eta_g] = 0 $. 
The factorization of the functional determinant carries over to the ghost determinant in the gauge-fixed partition function (cf. Eq.~\eqref{eq:generic_gf_partition_function}), which can be written as 
	\begin{equation} \label{eq:factorized_Z}
	Z= \int \mathcal{D} \eta\, \delta\big(G^\alpha[\eta] \big) \, \mathrm{Det} \big( G\ud{\alpha}{\!,I}[ \eta] D\ud{I}{\beta}[ \eta] \big)  \,\delta \big(G^i[\eta]\big) \mathrm{Det} \big( G\ud{i}{\!,I}[ \eta] D\ud{I}{j}[ \eta] \big) \, e^{iS[\eta]}.
	\end{equation}
This observation is due to Weinberg~\cite{Weinberg:1980wa}, who observed that since $ G^\alpha $ can be chosen to be independent of heavy fields, the factorization is ideally suited for integrating out the heavy DOFs and obtain the low-energy EFT.

Unfortunately, it is not possible to gauge fix to a constant $ \theta $ through the replacement $ G^i \to G^i - \theta^i $ to allow for the Gaussian weight gauge-fixed partition function~\eqref{eq:Gaussian_weight_gf_z}: this is incompatible with the covariance of $ G^i $ under $ H $~\eqref{eq:heavy_gf_H-invariance}. To mitigate the issue, Ferrari~\cite{Ferrari:2013aza} proposed the introduction of a dynamical auxiliary field $ \rho^{A} $ transforming in the adjoint of $ G $. It is included in a new term in the action,
	\begin{equation}
	S[\eta] \longrightarrow S[\eta] - \dfrac{1}{2\zeta} \int_x \brakets{\rho,\, \rho},
	\end{equation} 
which leaves the partition function~\eqref{eq:generic_partition_function} unchanged up to an irrelevant normalization factor:
	\begin{equation} 
	Z = \int \mathcal{D} \eta\, e^{iS[\eta]} = \int \mathcal{D} \eta\, \mathcal{D} \rho\, \exp\! \left[ iS[\eta] - \frac{i}{2\zeta} \int_x  \brakets{ \rho,\, \rho } \right]. 
	\end{equation} 
This choice introduces $ \zeta $ as an arbitrary gauge parameter for the $ G/H $ fixing. 
Next, we introduce the decomposition $ L\ud{A}{B}\rho^B = (\rho^\alpha,\, \rho^i) $. 
The inclusion of the auxiliary field allows for implementing the $ H $-covariant gauge-fixing condition
	\begin{equation}
	G^i \longrightarrow \mathcal{G}^i[\eta,\, \rho]= G^i[\eta] - \rho^i,
	\end{equation}
which transforms according to
	\begin{align} 
	\delta^H_\alpha \mathcal{G}^i[\eta,\,\rho] &=  G\ud{i}{\!,I}[\eta] D\ud{I}{\alpha}[\eta] \alpha^{\alpha} + f\ud{i}{\alpha j} \rho^j \alpha^{\alpha} = - f\ud{i}{\alpha j}  \mathcal{G}^{j}[\eta,\, \rho] \alpha^{\alpha},\\
	\delta^{G/H}_\alpha \mathcal{G}^i[\eta,\,\rho] &= ( G\ud{i}{\!,I}[\eta] D\ud{I}{j}[\eta] + f\ud{i}{jk} \rho^k + f\ud{i}{j \alpha} \rho^\alpha ) \alpha^j 
	\end{align}
for infinitesimal transformations in $ H $ and $ G/H $ respectively. 
The auxiliary field is included \emph{only} with the gauge-fixing condition for $ G/H $, not with the one for the unbroken group $ H $. 

The $ \rho^i $ integral in the partition function plays the role of the usual Gaussian weight integration. Meanwhile, the Gaussian $ \rho^\alpha $ integral is performed by completing the square. The result is that the partition function~\eqref{eq:factorized_Z} integrated over the auxiliary $ \rho $ field evaluates to
	\begin{equation} \label{eq:partially_fixed_Z}
	Z = \int \mathcal{D} \eta\,  \mathcal{D} \mathbf{u} \, \delta\big(G^\alpha[\eta] \big) \, \mathrm{Det} \big( G\ud{\alpha}{\!,I}[ \eta] D\ud{I}{\beta}[ \eta] \big)
	\exp \! \Big[ i \Big( S[\eta] + S^{G/H}_\mathrm{fix}[\eta,\, \mathbf{u}] \Big) \Big],
	\end{equation}
where $ \mathbf{u} = (u^i,\, \overline{u}_i)  $ are the ghost and antighost fields associated with $ V^i_\mu $. The partial gauge fixing term is 
	\begin{equation} \label{eq:G/H_fixing_term}
	S^{G/H}_\mathrm{fix}[\eta,\, \mathbf{u}] = -\! \int_x \bigg( \frac{1}{2\zeta} G_i[\eta] G^i[\eta] 
	+ \overline{u}_{i} \big( G\ud{i}{\! ,I}[\eta] D\ud{I}{j}[\eta] + f\ud{i}{jk} G^k[\eta] \big) u^j - \frac{\zeta}{2} \hat{a}^{\alpha\beta} f\ud{i}{j\alpha} f\ud{k}{\ell\beta} \overline{u}_{i} u^j \overline{u}_{k} u^\ell\bigg).
	\end{equation}
By construction $ S^{G/H}_\mathrm{fix} $ is invariant under $ H $ transformations, taking also the ghost fields $ \mathbf{u} $ to transform in the representation of the massive vectors. The price of this invariance is that the ghost action contains new quadratic and quartic ghost interactions~\cite{Ferrari:2013aza}. Obviously, the quartic interaction plays a role starting only at two-loop order, as there are no external ghost fields. 
We can choose the gauge of the remnant group independent of the partial fixing $ S^{G/H}_\mathrm{fix} $, which, as alluded to, turns out to be very convenient in EFT matching computations.

\subsection{Partially fixed background field gauge} \label{sec:partial_BF_gauge}
The closure of the remnant group $ H $ ensures that its gauge-fixing can be chosen entirely independent of the heavy vectors and effectively factorizes from the partial fixing of $ G/H $. 
We now combine the partial gauge-fixing of $ G/H $ with a BF gauge for $ H $ in order to construct an effective action that is explicitly invariant under $ H $.  Thus, we choose (cf. Eq.~\eqref{eq:BF_gauge_class})
	\begin{equation}
	G^\alpha[\eta,\, \Theta] = F\ud{\alpha}{I}[\Theta] (\eta - \Theta)^I,
	\end{equation}
where 
	\begin{equation} 
	\big( D\ud{J}{\beta}[\Theta] \alpha^\beta \big) F\ud{\alpha}{I,J}[\Theta] X^I +  F\ud{\alpha}{J}[\Theta] (D\ud{J}{\beta,I} \alpha^\beta X^I)= - f\ud{\alpha}{\beta \gamma} \alpha^\beta F\ud{\gamma}{I}[\Theta] X^I 
	\end{equation}
for any $ X^I $ and $ \alpha \in \mathfrak{h} $. The BFM vacuum functional of the partially gauge-fixed partition function~\eqref{eq:partially_fixed_Z}, constructed as in Eq.~\eqref{eq:BFM_W-tilde}, is given by 
	\begin{equation}
	\widetilde{\mathcal{W}}[J,\, \overline{\eta}] = - i \log \! \int \mathcal{D} \eta \, \mathcal{D} \mathbf{c} \, \mathcal{D} \mathbf{u} \, \exp \! \bigg[ i \bigg( S[\eta + \overline{\eta}] +  S^H_\mathrm{fix}[\eta + \overline{\eta},\, \mathbf{c}, \, \overline{\eta}] + S^{G/H}_\mathrm{fix}[\eta+ \overline{\eta},\, \mathbf{u}] + \int_x J_I \eta^I \bigg) \bigg],
	\end{equation}
where the gauge-fixing term of the stability group is 
	\begin{multline} \label{eq:stability_gauge-fixing_action}
	S_\mathrm{fix}^H[\eta,\, \textbf{c},\, \Theta]
	= - \!\int_x \bigg( \dfrac{1}{2\xi} \big\langle F_{I}[\Theta] (\eta - \Theta)^I,\, F_J[\Theta] (\eta  -\Theta)^J \big \rangle_{\mathfrak{h}} + \overline{c}_{\alpha} F\ud{\alpha}{I}[\Theta] D\ud{I}{\beta}[\eta]\, c^\beta	\bigg)
	\end{multline}
and $ \mathbf{c} = (c^\alpha,\, \overline{c}_\alpha) $ are the ghost fields associated with the massless $ B^\alpha_\mu $ fields.

The gauge-fixing of $ H $ is similar to that of the BF gauge of an unbroken gauge group. From the discussion in Section~\ref{sec:gauge-invariant_eff_action}, it follows that the gauge-fixing term $ S^H_\mathrm{fix}[\eta + \overline{\eta}, \, \mathbf{c},\, \overline{\eta}] $ is invariant w.r.t. background gauge transformations of the remnant group, which acts as  
	\begin{align}
	\overline{\delta}^H_\alpha \overline{\eta}^I &= D\ud{I}{\alpha}[\overline{\eta}] \alpha^\alpha, & 
	\overline{\delta}^H_\alpha \eta^I &= D\ud{I}{\alpha,J} \eta^J \alpha^\alpha,  \nonumber \\
	\overline{\delta}^H_\alpha c^\alpha &= - f\ud{\alpha}{\beta \gamma }c^\gamma \alpha^\beta, &
	\overline{\delta}^H_\alpha \overline{c}_\alpha &= f\ud{\gamma}{\beta \alpha} \overline{c}_\gamma \alpha^\beta.
	\end{align}
With the invariance of $ S $ and $ S^{G/H}_\mathrm{fix} $ under $ H $ transformations, this ensures that the vacuum functional $ \widetilde{\mathcal{W}}[J,\, \overline{\eta}] $ is invariant under background gauge transformations of the remnant group. The same applies to the gauge-invariant effective action $ \overline{\Gamma}[\overline{\eta}] $ associated with $ \widetilde{\mathcal{W}} $ (defined as in Eq.~\eqref{eq:gauge-invariant_effective_action}): in our construction, $ \overline{\Gamma}[\overline{\eta}] $ is also invariant under $ H $ background gauge transformations.

The $ H $ invariance of $ \overline{\Gamma}[\overline{\eta}] $ results in the familiar singularities of  2-point functions of $ \overline{B}^\alpha_\mu $, which prevents the direct construction of the connected Green's functions. Once again, we include a gauge-fixing term for the background fields to the effective action: 
	\begin{equation} \label{eq:part-fixed_BF_eff_action}
	\Gamma_\sscript{BF}[\overline{\eta}] = \overline{\Gamma}[\overline{\eta}] + S^H_\mathrm{bg.}[\overline{\eta}], \qquad 
	S^H_\mathrm{bg.}[\overline{\eta}] = \int_x \big\langle \mathcal{F}_I[\overline{\eta}] \overline{\eta}^I ,\, \mathcal{F}_J[\overline{\eta}] \overline{\eta}^J \big\rangle_{\mathfrak{h}},
	\end{equation}    
where $ \mathcal{F}\ud{\alpha}{I} $ is a gauge-fixing operator that can be chosen independently of $ F\ud{\alpha}{I} $ and can be chosen to be independent of $ \overline{\eta} $. $ \Gamma_\sscript{BF} $ in the partially fixed BF gauge can then be used to compute the generating functional of the connected Green's functions $ \mathcal{W}_{\sscript{BF}}[\overline{J}] $, as per Eq.~\eqref{eq:W_BF_definition}.  

Following an approach parallel to the one outlined in Section~\ref{sec:one-loop_eff_action}, we find that the partially invariant effective action up to one-loop order is given by   
	\begin{equation} \label{eq:part-fixed_invariant_eff_action}
	\overline{\Gamma}[\overline{\eta}] = S[\overline{\eta}] + S_\mathrm{fix}^{G/H}[\overline{\eta}, \, \mathbf{u} = 0] + \dfrac{i}{2} \mathrm{sTr} \log \mathcal{Q}^{\mathrm{all}}[\overline{\eta},\, \overline{\eta}].
	\end{equation}
In this case, the full, gauge-fixed action of the theory is given by 
	\begin{equation}
	S^{\mathrm{all}}[\eta, \, \mathbf{c},\, \mathbf{u},\, \Theta] = S[\eta] + S_\mathrm{fix}^{H}[\eta, \, \mathbf{c},\, \Theta] + S_\mathrm{fix}^{G/H}[\eta, \, \mathbf{u}],
	\end{equation} 
leading to the fluctuation operator 
	\begin{equation} 
	\mathcal{Q}^{\mathrm{all}}_{\mathbf{IJ}}[\underline{\eta},\, \Theta] = \dfrac{\delta^2 S^{\mathrm{all}}}{\delta \Omega^\mathbf{I} \delta \Omega^\mathbf{J}}[\underline{\eta},\, \mathbf{c}= 0,\, \mathbf{u}=0,\, \Theta], \qquad \Omega^\mathbf{I} = (\eta^I,\, \mathbf{c},\, \mathbf{u}).
	\end{equation}
The tree-level inclusion of the fixing-term $ S_\mathrm{fix}^{G/H}[\overline{\eta}, \, \mathbf{u} = 0] $ removes the singularities associated with the heavy vectors, and no background gauge-fixing term is required for these. The presence of this term also at tree level ensures that integrating out the heavy vectors at tree level is done with the selfsame Feynman rules that occur in the UV loops, ensuring the EFT--UV loop correspondence for hard-region matching.

\subsection{Choice of gauge-fixing conditions} \label{sec:part_fixed_gf_conditions}
Until now, the construction of the partially fixed BF gauge action has been completely generic. In practical calculation, we can specify a concrete gauge-fixing condition that gives convenient Feynman rules/fluctuation operators.

\subsubsection{Partial gauge-fixing condition for the broken directions}
We propose using the original gauge-fixing condition~\cite{Weinberg:1980wa} as a practical choice for $ G^i $. It is given by
	\begin{equation} \label{eq:gf_condition_G/H}
	G^i[\eta] = d^\mu V_\mu^i - \zeta f\ud{i}{a} \varphi^a,
	\end{equation}
which mimics the usual $ R_\xi $ gauge-fixing condition but using covariant derivatives of the unbroken group $ H $. $ \zeta $ is the arbitrary gauge fixing parameter of the gauge-fixing terms~\eqref{eq:G/H_fixing_term}. With the embedding of the would-be Goldstone Bosons (GBs) $ \chi^i $ in $ \varphi^a $, presented in Appendix~\ref{sec:GBs}, the condition is cast as
	\begin{equation}
	G^i[\eta] = d^\mu V_\mu^i - \zeta M_i \chi^i.
	\end{equation}
Thus, the gauge-fixing Lagrangian for the massive vectors is 
	\begin{equation} \label{eq:part_fixed_vectors}
	\L^{G/H}_\mathrm{vec.}[\eta] = - \frac{1}{2\zeta} G_i[\eta] G^i[\eta] = -\dfrac{1}{2\zeta} (d_\mu V^\mu_i) (d^\nu V^i_{\nu}) + M_i \chi_i d^\mu V_\mu^i - \dfrac{\zeta}{2} M_{i}^2 \chi_i \chi^i.
	\end{equation}
The kinetic mixing between massive vectors and GBs cancels the mixing terms from the scalar kinetic term by construction (in analogy to the $ R_\xi $-gauges). Clearly, the gauge~\eqref{eq:part_fixed_vectors} reproduces the propagators of the usual $ R_\xi $-gauges.\footnote{E.g., $ \zeta=1 $ corresponds to the Feynman gauge.} 

The fields relevant to the gauge-fixing condition are $ \eta^I = (B^\alpha_\mu,\, V^i_\mu,\, \varphi^{\prime a}) $. By performing a gauge variation in the direction of the broken algebra, we determine that 
	\begin{equation}
	D\ud{I}{j}[\eta] = \big(f\ud{\alpha}{kj} V^k_\mu ,\; \delta\ud{i}{j} d_\mu  + f\ud{i}{kj} V^k_\mu ,\; i x^a_{jb} \varphi^{\prime b} \big).
	\end{equation}
As $ \varphi' = \varphi + v $, the gauge transformation mixes the VEV with the quantum field. 
The derivative of the gauge-fixing function is given by 
	\begin{equation}
	G\ud{i}{\!,J}[\eta] = \big(f\ud{i}{\beta k} V^{k}_\mu,\; \delta\ud{i}{j} d_\mu ,\; - \zeta f\ud{i}{b} \big).
	\end{equation}
We can now derive the ghost term corresponding to the gauge-fixing condition:
	\begin{equation} \label{eq:part_fixed_ghosts}
	\begin{split}
	\L^{G/H}_\mathrm{gh.}[\eta, \mathbf{u}] = &\,- \overline{u}_i (G\ud{i}{\!,I} D\ud{I}{j} + f\ud{i}{jk} G^k) u^j + \frac{\zeta}{2} \hat{a}^{\alpha\beta} f\ud{i}{j\alpha} f\ud{k}{\ell\beta} \overline{u}_{i} u^j \overline{u}_{k} u^\ell\\
	=&\, -\overline{u}_i (d^2 + \zeta M_i^2) u^i +\overline{u}_i \big( f\ud{i}{jk} V^k_\mu d^\mu + f\ud{i}{k \alpha} f\ud{\alpha}{\ell j} V^{k\mu} V^{\ell}_\mu \big) u^j\\
	&\,+ \zeta \overline{u}_i \big( i f\ud{i}{a} x^a_{jb} \varphi^b + f\ud{i}{jk} M_k \chi^k \big) u^j 
	+ \frac{\zeta}{2} \hat{a}^{\alpha\beta} f\ud{i}{j\alpha} f\ud{k}{\ell\beta} \overline{u}_{i} u^j\,  \overline{u}_{k} u^\ell.
	\end{split}
	\end{equation}
These are all the terms that appear in the gauge-fixing action $ S^{G/H}[\eta,\, \mathbf{u}] $. In the gauge-invariant effective action, they appear with a background field, through $ \eta \to \eta + \overline{\eta} $, but this does not provide any additional complication: the Feynman rules from $ S^{G/H} $ are identical for the quantum and background fields. A practical example of how to derive these gauge-fixing terms from the broken algebra is shown in Appendix~\ref{app:toy_model}.

\subsubsection{BF Gauge-fixing condition for the unbroken direction}
For the BF gauge-fixing condition of the unbroken group, we take the standard choice for an unbroken gauge group and let
	\begin{equation} \label{eq:gf_condition_H}
	G_H^{\alpha}[\eta,\, \Theta] = F^\alpha_{H\,I}[\Theta] (\eta - \Theta)^I = \big( \delta\ud{\alpha}{\gamma} \partial^\mu + f\ud{\alpha}{\beta \gamma} \Theta^{\beta \mu}\big) (B- \Theta)_\mu^{\gamma}.
	\end{equation}
In the BF gauges with $ \Theta = \overline{\eta} $, we obtain the familiar 
	\begin{equation}
	G_H^{\alpha}[\eta+ \overline{\eta},\, \overline{\eta}] = F^\alpha_{H\, I}[\overline{\eta}] \eta^I = \overline{d}^\mu B^\alpha_\mu,
	\end{equation}
which is covariant under the background gauge transformation of the stability group. Here $ \overline{d}_\mu \equiv \partial_\mu - i \overline{B}_\mu $ is used for the background field covariant derivative w.r.t. the unbroken group. 
The resulting gauge-fixing term for the massless vectors is 
	\begin{equation} \label{eq:BF_vectors}
	\L^{H}_{\mathrm{vec.}}[\eta + \overline{\eta},\, \overline{\eta}] = - \frac{1}{2\xi} \brakets{G_H,\, G_H} = -\dfrac{1}{2\xi} \hat{a}^{\eminus 1}_{\alpha \beta} \overline{d}^\mu B^\alpha_\mu\, \overline{d}^\nu B^\beta_\nu.
	\end{equation}
The ghost term is given by 
	\begin{equation} \label{eq:BF_ghosts}
	\L_\mathrm{gh.}^{H}[\eta + \overline{\eta},\, \mathbf{c},\, \overline{\eta}] = - \overline{c}_\alpha F^{\alpha}_{H\,I}[\overline{\eta}]  D\ud{I}{\beta}[\eta+ \overline{\eta}] c^{\beta} = - \overline{c}_\alpha \overline{d}^\mu \big( \overline{d}_\mu c^\alpha + f\ud{\alpha}{\beta \gamma} B^\beta_\mu c^\gamma\big), 
	\end{equation}
where there is a massless quantum gauge field that is not in a covariant derivative, consistent with the breaking of quantum gauge invariance. This is nothing but an ordinary $ R_\xi $-like gauge-fixing condition for $ H $ in the background field gauge. 

The two gauge-fixing conditions Eqs.~\eqref{eq:gf_condition_G/H} and~\eqref{eq:gf_condition_H} both result in propagators for the quantum vector fields similar to that of the $ R_\xi $-gauges. They are expected to result in a convenient class of gauges for matching computations, as they allow for the calculation of an effective action that is explicitly gauge invariant w.r.t. the unbroken group $ H $. The reader may compare the gauge-fixing terms for the partially fixed BF gauge with those of the ordinary BF gauge in Appendix~\ref{app:broken_phase_BF-gauge} to appreciate their relative simplicity. 
The subsequent discussion of the matching formula is kept generic and does not rely on these particular choices for the gauge-fixing conditions.

\section{EFT Matching In Gauge Theories} \label{sec:matching}

UV theories usually come with one or more mass thresholds, below which heavy fields become non-dynamical and can be dropped from the description both for simplicity and to use RG methods to resum the large logarithm common to perturbation theory. Given a UV theory $ S[\eta] $\footnote{We omit the subscript `UV' everywhere to improve legibility} with a threshold $ \Lambda $, we decompose the field as $ \eta^I= (\Phi^x,\, \phi^a) $, where $ \Phi $ denote the collection of heavy fields with masses $ M \gtrsim \Lambda $ and $ \phi $ is the collection of light (or massless) fields with masses $ m \ll \Lambda $. The role of EFT matching is to determine a local EFT action $ S_\EFT[\phi] $ that provides a valid description of physics for energies below $ \Lambda $ with only light dynamical fields.

Off-shell matching turns out to be very convenient for practical calculations of $ S_\EFT $. It is based on a strong matching condition equating the generating functionals for the UV and EFT theories, a sufficient but not necessary condition for the EFT to reproduce the physics of the UV theory. An immediate concern is that the gauges of the two theories should be chosen to be compatible to have any chance of enforcing the strong matching condition: given a particular gauge-fixing condition for the UV theory, there will be at most one choice for the EFT that ensure off-shell equality; however, we are unaware of anything that guarantees the existence of \emph{any} EFT gauge that satisfy the condition. Our first objective is to show that matching with the strong condition is possible (at one-loop order) in unbroken gauge theories using a BF gauge. Establishing the correspondence of EFT loops and soft-region UV loops enables the usual one-loop matching formula and the construction of a local EFT action. 

In the SSB scenario, it is not possible to choose the same gauge-fixing condition in the UV theory and the EFT. Clearly, the gauge-invariant effective actions cannot be equated directly: the EFT effective action satisfies an $ H\subseteq G $ background gauge symmetry, while the UV action is symmetric under the full symmetry  $ G $. With fixing of the background fields this can be resolved, but it is then unclear if it is possible to establish the one-to-one loop correspondence. The partially fixed BF gauge offers a solution because it reduces the background gauge invariance of the UV theory to $ H $, which may be gauge-fixed in an identical manner in the EFT. Thus, we may establish the correspondence in this gauge.

\subsection{EFT matching with the BF gauge in unbroken gauge theories}
We begin by considering the case of EFT matching in an unbroken gauge theory with gauge group $ G $. The gauge fields are massless, meaning that $ A^A_\mu \subset \phi^a $. The ghost fields are massless too, so they appear as dynamical DOFs in both the UV and EFT theories. We indicate the set of light quantum fields with $ \psi^\mathbf{a} = (\phi^a,\, \omega^A,\, \overline{\omega}_A) $. 

The generating functional for the connected Green's functions in a BF gauge, $ \mathcal{W}_\sscript{BF}[\overline{J}] $, contains the physics of the theory in that it is sufficient to generate the $ S $-matrix elements. 
We enforce the strong matching condition 
	\begin{equation}
	\mathcal{W}_\sscript{BF}^{\EFT}[\overline{J}^\phi] = \mathcal{W}_\sscript{BF}[\overline{J}^\Phi= 0,\, \overline{J}^\phi]
	\end{equation}
at the level of connected Green's functions; no heavy fields go on shell at low energies and accordingly, they remain unsourced. Here $ \overline{J} = (\overline{J}^\Phi,\, \overline{J}^\phi) $ are the BF sources associated with $ \overline{\Phi} $ and $ \overline{\phi} $, respectively.
The strong matching condition is sufficient to ensure that the EFT reproduces the low-energy physics of the UV theory (the $ S $-matrix can be extracted from $ \mathcal{W}_\sscript{BF} $), and it is understood at the level of power series in $ E/\Lambda $ and can be truncated as convenient in practical calculations. From the defining Legendre transformation~\eqref{eq:W_BF_definition}, we obtain the equivalent matching condition 
	\begin{equation}  \label{eq:matching_condition_BF}
	\Gamma_\sscript{BF}^{\EFT}[\overline{\phi}] = \Gamma_\sscript{BF}\big[\overline{\Phi}[\overline{\phi}],\, \overline{\phi} \big], \qquad 
	0= \dfrac{\delta \Gamma_\sscript{BF}}{\delta \overline{\Phi}} \big[\overline{\Phi}[\overline{\phi}],\, \overline{\phi} \big] 
	\end{equation}
for the BF gauge effective actions. The heavy background fields are not independent degrees of freedom but are given as the solutions to the quantum field EOMs in the presence of the light background fields. It is sufficient to use the tree-level solutions for the heavy field EOM in one-loop matching, and we will use $ \overline{\eta} = \big(\overline{\Phi}[\overline{\phi}],\, \overline{\phi} \big) $ as a useful short-hand notation in the rest of this section.

\subsubsection{The EFT effective action}\label{sec:BF_EFT_eff_action}
The aim of EFT matching is to determine a gauge-invariant (non-fixed) EFT action $ S_\EFT[\phi] $ that reproduces the low-energy physics of the UV theory. Clearly, the matching condition~\eqref{eq:matching_condition_BF} entails gauge-fixing the EFT at intermediate steps but ideally, we should be able to disentangle the fixing from $ S_\EFT $ in the end. 
Given a gauge-fixing operator $ F' $ and a corresponding gauge-fixing term $  S^G_\mathrm{fix}[\phi+ \overline{\phi} ,\, \boldsymbol{\omega},\, \overline{\phi};\, F'] $, we can construct the BF effective action of the EFT:
	\begin{equation}
	\Gamma_\sscript{BF}^\EFT[\overline{\phi};\, F'] = \overline{\Gamma}_\EFT[ \overline{\phi};\, F'] + S_\mathrm{bg.}[\overline{\phi};\, \mathcal{F}'].
	\end{equation}
In the notation of Section \ref{sec:gauge-invariant_eff_action}, the background gauge fields are fixed with the operator $ \mathcal{F}' $.
We have explicitly included the dependence on the gauge-fixing operators for future clarity.
The DOFs relevant to the EFT loops are $ \psi^\mathbf{a} $, so it follows from Eq.~\eqref{eq:one-loop_GI_effecitve_action} that the effective action up to one-loop order is 
	\begin{equation} \label{eq:EFT_eff_action}
	\Gamma_\sscript{BF}^\EFT[\overline{\phi};\, F'] = S_\EFT[\overline{\phi}] + S_\mathrm{bg.}[\overline{\phi};\, \mathcal{F}'] + \dfrac{i}{2} \mathrm{sTr} \log \mathcal{Q}^{\mathrm{all} }_{\EFT\, \mathbf{ab}}[ \overline{\phi},\, \overline{\phi};\, F'],
	\end{equation}
where the EFT fluctuation operator takes the form 
	\begin{equation}
	\mathcal{Q}^{\mathrm{all}}_{\EFT\, \mathbf{ab}}[\overline{\phi},\, \overline{\phi};\, F'] = \mathcal{Q}_{\EFT\, \mathbf{ab}}[\overline{\phi}] + \mathcal{Q}^{G\text{-fix}}_{\mathbf{ab}}[\overline{\phi},\, \overline{\phi};\, F'].
	\end{equation}
The first term in the fluctuation operator is due to $ S_\EFT $ and is singular. The addition of the second term, due to $ S^G_\mathrm{fix} $, lifts these singularities. As the matching is performed order by order in the loop expansion, it will be useful to write 
	\begin{equation}
	S_\EFT =  S_\EFT^{(0)} + S_\EFT^{(1)} + \ldots ,
	\end{equation}
where $ S_\EFT^{(\ell)} $ denotes the $ \ell $-loop contribution to $ S_\EFT $.

\subsubsection{The UV effective action}
Since all the gauge fields are light, the gauge-fixing of the background fields can be taken to be independent of the heavy fields. Thus, the BF effective action of the UV theory is 
	\begin{equation}
	\Gamma_\sscript{BF}[\overline{\eta};\, F] = \overline{\Gamma}[ \overline{\eta};\, F] + S_\mathrm{bg.}[\overline{\phi};\, \mathcal{F}].
	\end{equation} 
Similarly, we choose a gauge-fixing operator for the quantum fields that does not involve any heavy fields, i.e.,\footnote{This is the case for, e.g., the standard choice for the background field gauge $ G^A[\eta + \overline{\eta}, \overline{\eta}] = \overline{D}^\mu \! A_\mu^A $.}
	\begin{equation}
	S_\mathrm{fix}^G[\eta + \overline{\eta},\, \boldsymbol{\omega},\, \Theta= \overline{\eta};\, F] = S^G_\mathrm{fix}[\phi+ \overline{\phi} ,\, \boldsymbol{\omega},\, \overline{\phi};\, F].
	\end{equation}
At this stage, we may think of the gauge-fixing operators of the UV action as independent from those of the EFT. 
From Eq.~\eqref{eq:one-loop_GI_effecitve_action}, it follows that the BF effective action up to one-loop order is 
	\begin{equation} \label{eq:UV_eff_action}
	\Gamma_\sscript{BF}[\overline{\eta};\, F] = S[\overline{\eta}] + S_\mathrm{bg.}[\overline{\phi};\, \mathcal{F}] + \dfrac{i }{2} \mathrm{sTr} \log \mathcal{Q}_{\mathbf{IJ}}^{\mathrm{all}}[\overline{\eta},\, \overline{\phi};\, F].
	\end{equation}
In terms of heavy and light components, $ (\Phi^x,\, \psi^\mathbf{a}) $, the fluctuation operator~\eqref{eq:fluct_op} decomposes as 
	\begin{equation}
	\mathcal{Q}^{\mathrm{all}}_{\mathbf{IJ}}[\overline{\eta},\, \overline{\eta};\, F] = 
	\mathcal{Q}_{\mathbf{IJ}}[\overline{\eta}] +\mathcal{Q}^{G}_{\mathbf{IJ}}[\overline{\phi},\, \overline{\phi};\, F] =
	\begin{pmatrix}
	\mathcal{Q}_{xy}[\overline{\eta}] & \mathcal{Q}_{x\mathbf{b}}[\overline{\eta}] \\
	\mathcal{Q}_{\mathbf{a}y}[\overline{\eta}] & \mathcal{Q}_{\mathbf{ab}}[\overline{\eta}] + \mathcal{Q}^{G}_{\mathbf{ab}}[\overline{\phi},\, \overline{\phi};\, F]
	\end{pmatrix}
	\end{equation}
with contributions from the UV action $ S $ and the gauge-fixing action $ S^{G}_\mathrm{fix} $, respectively. That $ S^{G}_\mathrm{fix} $ only contributes to the light block is key to establishing the matching formula. 

For later use, it will prove useful to introduce the matrix\footnote{To be explicit, by $ \mathcal{Q}^{\eminus 1}_{xy} $ we indicate the inverse of $ \mathcal{Q}_{xy} $, \emph{not} the heavy, $ xy $-block of $ \mathcal{Q}^{\eminus 1}_{\mathbf{IJ}} $.}
	\begin{equation}
	V\ud{\mathbf{I}}{\mathbf{J}}[\overline{\eta}] = \begin{pmatrix}
	\delta\ud{x}{y} & - \mathcal{Q}^{\eminus 1}_{xz} \mathcal{Q}_{z\mathbf{b}} \\ 0 & \delta\ud{\mathbf{a}}{\mathbf{b}}
	\end{pmatrix},
	\end{equation}
which block-diagonalizes the fluctuation operator associated with the unfixed UV action $ S $. We have 
	\begin{equation} \label{eq:X-operator_unbroken}
	X_{\mathbf{IJ}} \equiv V\ud{\mathbf{K}}{\mathbf{I}} \mathcal{Q}_{\mathbf{KL}} V\ud{\mathbf{L}}{\mathbf{J}} = \begin{pmatrix}
	\mathcal{Q}_{xy} & 0 \\ 0 & \mathcal{Q}_{\mathbf{ab}} - \mathcal{Q}_{\mathbf{a}z} \mathcal{Q}_{zu}^{\eminus 1} \mathcal{Q}_{u\mathbf{b}}
	\end{pmatrix}.
	\end{equation}
This diagonalization is possible because the gauge fields are light, and $ \mathcal{Q}_{xy} $, therefore, non-singular. Although the lack of fixing in the gauge kinetic term and the overall absence of ghost terms in $ S $ means that $ \mathcal{Q}_{\mathbf{ab}} $ is singular by itself, this is not an issue for the block diagonalization.

\subsubsection{One-loop matching} \label{sec:unbroken_one-loop_matching}
At tree level, the matching condition~\eqref{eq:matching_condition_BF} reduces to 
	\begin{equation}
	S^{(0)}_\EFT[\overline{\phi}] + S_\mathrm{bg.}[\overline{\phi};\, \mathcal{F}'] = S\big[\overline{\Phi}[\overline{\phi}],\, \overline{\phi} \big] + S_\mathrm{bg.}[\overline{\phi};\, \mathcal{F}].
	\end{equation}
As one might expect, it is indeed convenient to choose $ \mathcal{F}' = \mathcal{F} $ at this stage; it is presumably even required in order to obtain a gauge-invariant $ S_\EFT[\phi] $. 
With this choice, the gauge-fixing for the background gauge fields drops out and can be ignored completely in the matching computation. These are the only terms that break background gauge invariance, so the rest of the computation is explicitly background gauge invariant, which is a great benefit in terms of simplicity.
We find the gauge-invariant EFT action 
	\begin{equation} \label{eq:matching_0_ord_unbroken}
	S^{(0)}_\EFT[\overline{\phi}] = S[\eta^s] \equiv S\big[\overline{\Phi}^\mathrm{s}[\overline{\phi}],\, \overline{\phi} \big],
	\qquad 0 = \dfrac{\delta S}{\delta \Phi^x}[\eta^\mathrm{s}] ,
	\end{equation} 
where the superscript ``s'' denotes the series expansion of the heavy field EOM in $ 1/\Lambda $. With no gauge fields among the heavy fields, the solution to the EOM is unique. Thus, we arrive at the expected result that tree-level matching in unbroken gauge theories can be done by solving the tree-level EOMs of the heavy fields (perturbatively in the $ 1/\Lambda $ expansion) and plugging it into the UV action; no gauge fixing is required in this case.

We will proceed as in Refs.~\cite{Zhang:2016pja,Fuentes-Martin:2016uol} to demonstrate that all soft-scale loops in the UV theory cancel against the EFT loops, which will enable us to derive a one-loop matching formula. First, we must determine the EFT fluctuation operator. By applying a $ \psi^\mathbf{a} $ derivative to the tree-level EOM satisfied by the heavy fields $ \overline{\Phi}^x[\phi] $, one finds that
	\begin{equation}
	\dfrac{\delta \overline{\Phi}^x[\phi]}{\delta \psi^\mathbf{a}} = - \mathcal{Q}_{xy}^{\eminus 1}[\overline{\Phi}[\phi],\, \phi] \mathcal{Q}_{y \mathbf{a}}[\overline{\Phi}[\phi],\, \phi].
	\end{equation}
With this result and a bit of algebra, one can then show that the soft-scale fluctuation operator of the EFT is given by 
	\begin{equation} \label{eq:EFT_fluct_unbroken}
	\mathcal{Q}_{\EFT\, \mathbf{ab}}[\overline{\phi}] = \dfrac{\delta^2 S_\EFT^{(0)}}{\delta \psi^\mathbf{a} \delta \psi^\mathbf{b}}[\overline{\phi}] = X_{\mathbf{ab}}[\eta^\mathrm{s}], \qquad \text{(in soft-scale loops)},
	\end{equation}
where $ X_{\mathbf{ab}} $ is defined as in Eq.~\eqref{eq:X-operator_unbroken}.	
Strictly speaking, $ \mathcal{Q}_{\EFT\, \mathbf{ab}} $ is a local differential operator since \eqref{eq:matching_0_ord_unbroken} is a series expansion in $ 1/\Lambda $. This is in contrast to $ X_{\mathbf{ab}} $, which involves the unexpanded $ \mathcal{Q}^{\eminus 1}_{xy} $. The equality~\eqref{eq:EFT_fluct_unbroken} is therefore valid, only for soft-scale loops, where the loop momentum is small compared to $ \Lambda $. This restriction is irrelevant, as all EFT loops are soft-scale: the hard region EFT loops are scaleless and vanish.
Thus, we find that the genuine loop contribution to the EFT effective action is given by 
	\begin{equation}
	\mathrm{sTr} \log \mathcal{Q}^{\mathrm{all}}_{\EFT\,\mathbf{ab}}[\overline{\phi},\, \overline{\phi};\, F'] 
	= \mathrm{sTr} \log\! \big( X_{\mathbf{ab}}[\eta^\mathrm{s}] +  \mathcal{Q}_{\mathbf{ab}}^{G}[\overline{\phi},\, \overline{\phi};\, F'] \big) \Big|_{\mathrm{soft}}.
	\end{equation}
Whereas $ X_{\mathbf{ab}} $ is singular, the gauge-fixing terms of $ \mathcal{Q}_{\mathbf{ab}}^{G} $ removes the singularities of ghosts and vectors.

The loop contributions of the UV effective action are split in a sum of the hard and the soft regions of the loop integrals with the method of regions~\cite{Beneke:1997zp,Jantzen:2011nz}.\footnote{
	The decomposition relies on a counterintuitive property of dimensionally regulated loop integrals. Any one-loop integral satisfies
	\begin{equation*}
	\int \dd^d k\, I(k) = \int \dd^d k\, I(k) \bigg|_{\mathrm{soft}} + \int \dd^d k\, I(k)\bigg|_{\mathrm{hard}},
	\end{equation*}
where the integrand $ I(k) $ is expanded around $ k\ll \Lambda $ in the soft region and $ k\gtrsim \Lambda $ in the hard region, but the integration region is over all of momentum space in both integrals. The identity holds to all orders, and the regions can be expanded up to any order required for the particular calculation. 
} 
In the soft region, the fluctuation operator is block-diagonalized with $ V\ud{\mathbf{I}}{\mathbf{J}} $, which we are free to do since $ \mathrm{sDet}\, V\ud{\mathbf{I}}{\mathbf{J}} = 1 $. We find that
	\begin{equation}
	\begin{split}
	\mathrm{sTr} \log \mathcal{Q}^{\mathrm{all}}_{\mathbf{IJ}}[\eta^\mathrm{s},\, \overline{\phi};\, F] = \mathrm{sTr} \log \mathcal{Q}^{\mathrm{all}}_{\mathbf{IJ}}[\eta^\mathrm{s},\, \overline{\phi};\, F] \Big|_\mathrm{hard} + \mathrm{sTr} \log\!  \big( X_{\mathbf{ab}}[\eta^s] + \mathcal{Q}_{\mathbf{ab}}^{G}[\overline{\phi},\, \overline{\phi};\, F] \big) \Big|_\mathrm{soft} ,
	\end{split}
	\end{equation} 
having also utilized that soft loops of $ X_{xy} =\mathcal{Q}_{xy} $ are scaleless, which makes them vanish in dimensional regularization. This central construction of Refs.\cite{Zhang:2016pja,Fuentes-Martin:2016uol} works because the block diagonalization works trivially on the gauge-fixing contribution to the fluctuation operator: $ V\ud{\mathbf{K}}{\mathbf{I}} \mathcal{Q}_{\mathbf{KL}}^{G} V\ud{\mathbf{L}}{\mathbf{J}} = \mathcal{Q}_{\mathbf{IJ}}^{G} $.

Plugging everything back into matching condition~\eqref{eq:matching_condition_BF} yields the one-loop order EFT action
	\begin{multline} \label{eq:incomplete_matching_formula}
	S_\EFT^{(1)}[\overline{\phi}] = \dfrac{i}{2} \Big[ \mathrm{sTr} \log \mathcal{Q}^{\mathrm{all}}_{\mathbf{IJ}}[\eta^\mathrm{s},\, \overline{\phi};\, F] \Big|_\mathrm{hard} 
	+ \mathrm{sTr} \log\!  \big( X_{\mathbf{ab}}[\eta^s] + \mathcal{Q}_{\mathbf{ab}}^{G}[\overline{\phi},\, \overline{\phi};\, F] \big) \Big|_\mathrm{soft} \\
	- \mathrm{sTr} \log\!  \big( X_{\mathbf{ab}}[\eta^s] + \mathcal{Q}_{\mathbf{ab}}^{G}[\overline{\phi},\, \overline{\phi};\, F'] \big) \Big|_\mathrm{soft}
	\Big].
	\end{multline}
We can ensure that the soft-region loops all cancel by simply choosing $ F' = F $, something that we may freely do to best evaluate $ S_\EFT $. We obtain a familiar form for the one-loop matching formula:
	\begin{equation} \label{eq:unbroken_matching_formula}
	S_\EFT^{(1)}[\overline{\phi};\, F] = \dfrac{i }{2} \mathrm{sTr} \log \mathcal{Q}^{\mathrm{all}}_{\mathbf{IJ}} [\eta^\mathrm{s},\, \overline{\phi};\, F] \Big|_\mathrm{hard},
	\end{equation}
where the hard scale loop is computed in a BF gauge. We note that $ S_\EFT^{(1)} $ may depend on~$ F $ (and for that matter the gauge parameter). However, $ S_\EFT $ should still reproduce the correct physics in all cases. A convenient corollary of the matching formula~\eqref{eq:unbroken_matching_formula} is that when using the BF gauge for matching calculations with unbroken gauge symmetries in the manner presented here, the ghost loops drop out: the ghosts are massless and result in scaleless loops in the hard region. The matching formula derived here demonstrates the validity of the common approach to matching in gauge unbroken gauge theories.

\subsection{EFT matching with the partially fixed BF gauge}

Spontaneous breaking of the gauge symmetry can complicate the EFT matching calculations; however, with the partially fixed BF gauge suggested in Section~\ref{sec:partial_BF_gauge}, many of the complexities vanish and the matching procedure is reminiscent of that outlined for the unbroken gauge group. 
We will assume a common scenario where the VEV of the scalar fields breaks the gauge group $ G \to H $ at the scale $ \Lambda $ and the gauge coupling constants are $ \mathcal{O}(1) $. Thus, the gauge fields $ B_\mu $ of the remnant group and their associated ghost fields $ \mathbf{c} $ are light (massless) compared to $ \Lambda $, whereas the massive vectors $ V_\mu $, their ghost fields $ \mathbf{u} $, and the associated would-be GBs $ \chi $ have masses comparable to $ \Lambda $ (in Feynman-like gauges) and decouple. The remaining scalars of the theory, whether or not they partake in the symmetry breaking, can in general be a mix of heavy and light states.   
In this situation the light DOFs are $ \psi^{\mathbf{a}} = (\phi^a,\, \mathbf{c}^{\textsf{a}}) $ and the heavy ones $ \Psi^{\mathbf{x}} = (\Phi^x,\, \mathbf{u}^\textsf{x}) $, where, e.g., $ \mathbf{c}^{\textsf{a}} = (c^\alpha,\, \overline{c}_\alpha) $. 
The gauge group of the EFT is the unbroken group $ H $, and all fields organize as irreducible representations of $ H $. 

The scenario described here, is very general covering many BSM situations. In many cases BSM theories will involve a sequence of VEVs each breaking the original symmetry to smaller and smaller remnant groups: $ G_1 \supseteq G_2 \supseteq \ldots  $. If there is a hierarchical separation of these scales, one would then perform sequential matching step from a $ G_n $ gauge group onto the unbroken phase of the corresponding remnant group $ G_{n+1} $ before RG running the new EFT down to the next breaking scale and repeating. If there is no hierarchy between consecutive breakings, it would be more appropriate to think of one big matching step rather than two consecutive ones. The main exception to our assumption is when we need the EFT to include the massive vectors because they are light compared to $ \Lambda $. This can happen for instance when integrating out a heavy Higgs field or if the gauge couplings are small, making the massive vectors light compared to the VEV. In such situations it may be more appropriate to describe the EFT with a non-linearly realized $ G $ symmetry~\cite{Dittmaier:1995cr,Dittmaier:2021fls} rather than the linearly realized remnant $ H $.

We will now derive a matching formula for the case of spontaneous breaking of the gauge group of the UV theory. The starting point is the same strong matching condition~\eqref{eq:matching_condition_BF}, as in the case of an unbroken gauge group; however, now we use the partially fixed BF gauge for the UV effective action $ \Gamma_\sscript{BF} $. The EFT effective action is also produced in the BF gauge, such that effectively both UV and EFT theories share the BF gauge for unbroken group $ H $.

\subsubsection{The EFT effective action}
The EFT action in the BF gauge is constructed entirely parallel to Section~\eqref{sec:BF_EFT_eff_action}; we merely repeat it here to make it explicit in the notation employed in the broken theory. 
Given a BF gauge gauge-fixing operator $ F'_H $ for the unbroken gauge group $ H $ and a corresponding gauge-fixing term $  S^H_\mathrm{fix}[\phi+ \overline{\phi} ,\, \mathbf{c},\, \overline{\phi};\, F'_H] $, the BF effective action of the EFT is
	\begin{equation} \label{eq:EFT_eff_action_H}
	\Gamma_\sscript{BF}^\EFT[\overline{\phi};\, F'_H] =  S_\EFT[\overline{\phi}] + S^H_\mathrm{bg.}[\overline{\phi};\, \mathcal{F}'_H] + \dfrac{i}{2} \mathrm{sTr} \log \mathcal{Q}^{\mathrm{full}}_\EFT[\overline{\phi},\, \overline{\phi};\, F_H'].
	\end{equation}
$ \mathcal{F}'_H $ is the gauge-fixing operator for the background fields. 
In this case, the fluctuation operator takes the form 
	\begin{equation}
	\mathcal{Q}^{\mathrm{all}}_{\EFT\, \mathbf{ab}}[\overline{\phi},\, \overline{\phi};\, F_H'] = \mathcal{Q}_{\EFT\, \mathbf{ab}}[\overline{\phi}] + \mathcal{Q}^{H}_{\mathbf{ab}}[\overline{\phi},\, \overline{\phi};\, F'_H].
	\end{equation}
The first term in the fluctuation operator is due to $ S_\EFT $, while the second is due to $ S^H_\mathrm{fix} $.

\subsubsection{The UV effective action}
The gauge fields associated with the unbroken group $ H $ are massless, and we use gauge-fixing conditions for the background fields and BF gauge for $ H $ that depend only on the light fields. Following Eqs.~\eqref{eq:part-fixed_BF_eff_action} and~\eqref{eq:part-fixed_invariant_eff_action}, the partially fixed BF effective action of the UV theory is then given by 
	\begin{equation} 
	\Gamma_\sscript{BF}[\overline{\eta}] =  S[\overline{\eta}] + S_\mathrm{fix}^{G/H}[\overline{\eta}, \, \mathbf{u} = 0;\, G] + S^H_\mathrm{bg.}[\overline{\phi};\, \mathcal{F}_H ]
	 + \dfrac{i}{2} \mathrm{sTr} \log \mathcal{Q}^{\mathrm{all}}[\overline{\eta},\, \overline{\phi};\, F_H,\, G],
	\end{equation}
where $ G^{i} $ is the gauge-fixing condition for the massive vectors associated to $ G/H $.
The fluctuation operator of the UV theory decomposes as 
	\begin{equation}
	\mathcal{Q}^{\mathrm{all}}_{\mathbf{IJ}}[\overline{\eta},\, \overline{\phi};\, F_H,\, G] = \mathcal{Q}_{\mathbf{IJ}}[\overline{\eta}] + \mathcal{Q}^{G/H}_{\mathbf{IJ}}[\overline{\eta};\, G] + \mathcal{Q}^{H}_{\mathbf{IJ}}[\overline{\phi},\, \overline{\phi};\, F_H],
	\end{equation}
with contributions from $ S $, $ S^{G/H}_{\mathrm{fix}} $, and $ S^H_\mathrm{fix} $, respectively.
The ghost number is conserved in the gauge-fixing terms, while there are no background ghost fields. Consequently, the fluctuation operator block-diagonalizes into a part with regular fields and a part with ghost fields. Schematically, we write
	\begin{equation}
	\mathcal{Q}^{\mathrm{all}}_{\mathbf{IJ}}[\overline{\eta},\, \overline{\phi};\, F_H,\, G] = 	\begin{pmatrix}
		\mathcal{Q}'_{xy} & \mathcal{Q}'_{x b}\\
		\mathcal{Q}'_{a y} & \mathcal{Q}'_{ab}+ \mathcal{Q}^{H}_{ab}
	\end{pmatrix} \oplus
	\begin{pmatrix}
		\mathcal{Q}^{G/H}_{\textsf{xy}} & 0 \\
		0 & \mathcal{Q}^{H}_{\textsf{ab}}
	\end{pmatrix},
	\end{equation}
where $ \mathcal{Q}' = \mathcal{Q} + \mathcal{Q}^{G/H} $. Note the distinction between sans-serif indices for the ghost DOFs and italic indices for the regular DOFs. The use of the partially fixed BF gauge has ensured that the ghost fluctuation operator is already block diagonalized and that the block corresponding to the light ghost field is determined entirely by the choice of BF gauge condition.

With the partial fixing term for the heavy fields, $ \mathcal{Q}'_{xy} $ is non-singular. This lets us introduce the matrix 
	\begin{equation}
	V\ud{\mathbf{I}}{\mathbf{J}}[\overline{\eta};\, G] = \begin{pmatrix}
		\delta\ud{x}{y} & - \mathcal{Q}^{\prime \eminus 1}_{xz} \mathcal{Q}'_{zb} \\ 0 & \delta\ud{a}{b}
	\end{pmatrix} \oplus \delta\ud{\textsf{I}}{\textsf{J}}
	\end{equation}
to block-diagonalize the partially fixed fluctuation operator. We define 
	\begin{equation} \label{eq:X-operator_part_fixed}
	X_{\mathbf{IJ}}
	\equiv V\ud{\mathbf{K}}{\mathbf{I}} \mathcal{Q}'_{\mathbf{KL}} V\ud{\mathbf{L}}{\mathbf{J}} = \begin{pmatrix}
		\mathcal{Q}'_{xy} & 0 \\ 0 & \mathcal{Q}_{ab}' - \mathcal{Q}'_{az} \mathcal{Q}^{\prime \eminus 1}_{zu} \mathcal{Q}'_{ub} 
	\end{pmatrix} \oplus
	\begin{pmatrix}
		\mathcal{Q}^{G/H}_{\textsf{xy}} & 0 \\
		0 & 0
	\end{pmatrix}
	\end{equation}
as the block diagonalized $ \mathcal{Q}' $. The same matrix, $ V $, block-diagonalizes the full fluctuation operator $ \mathcal{Q}^\mathrm{all} $ too.

\subsubsection{One-loop matching}
Let us now use the matching condition~\eqref{eq:matching_condition_BF} to derive a formula for the EFT action. At tree level, the condition reduces to 
	\begin{equation}
	S^{(0)}_\EFT[\overline{\phi}] + S_\mathrm{bg.}[\overline{\phi};\, \mathcal{F}'_H] = S[ \overline{\eta} ]+ S_\mathrm{fix}^{G/H}[\overline{\eta}, \, \mathbf{u} = 0;\, G] + S_\mathrm{bg.}[\overline{\phi};\, \mathcal{F}_H].
	\end{equation}
Again the choice for the BF gauge-fixing operator falls on $ \mathcal{F}'_H = \mathcal{F}_H $ in order to obtain the $ H $ gauge-invariant EFT action 
	\begin{equation} \label{eq:matching_0_ord_part_fixed}
	S^{(0)}_\EFT[\overline{\phi}] = S'[\eta^s] ,
	\qquad S'[\eta] = S[\eta] + S^{G/H}_\mathrm{fix}[\eta,\, 0;\, G].
	\end{equation} 
The heavy background fields are solutions of their EOMs in the presence of the light fields:
	\begin{equation}
	0 = \dfrac{\delta S'}{\delta \Phi^x}[\eta^\mathrm{s}] 
	\end{equation}
understood to be a power series in $ 1/\Lambda $ and with higher order corrections necessary for matching beyond one-loop order. As one might anticipate, the gauge-fixing condition for the heavy vectors is added to the invariant UV action, which ensures a unique solution to the heavy field EOMs. 

In analogy to Eq.~\eqref{eq:EFT_fluct_unbroken}, the tree-level matching condition for the EFT action is used to determine the fluctuation operator of the EFT: 
	\begin{equation} 
	\mathcal{Q}_{\EFT\, ab}[\overline{\phi};\, G] = \dfrac{\delta^2 S_\EFT^{(0)}}{\delta \phi_a \delta \phi_b}[\overline{\phi}] = X_{ab}[\eta^\mathrm{s}; \,G], \qquad \text{(in soft-scale loops)},
	\end{equation}
where $ X_{ab} $ is defined in Eq.~\eqref{eq:X-operator_part_fixed}. The soft-scale correspondence between EFT and UV fluctuation operators is due to the partial-fixing terms of the heavy vectors being the same for the UV loops and the tree-level matching condition for the UV.
We proceed as in Section~\ref{sec:unbroken_one-loop_matching} to decompose the UV loops into a hard and a soft part while block-diagonalizing the fluctuation operator with $ V\ud{\mathbf{I}}{\mathbf{J}} $. The corresponding version of Eq.~\eqref{eq:incomplete_matching_formula} now takes the form
	\begin{multline} \label{eq:part_fixed_soft_cancellation}
	S_\EFT^{(1)}[\overline{\phi}] = \dfrac{i}{2} \Big[ \mathrm{sTr} \log \mathcal{Q}^{\mathrm{all}}_{\mathbf{IJ}}[\eta^\mathrm{s},\, \overline{\phi};\, F_H,\, G] \Big|_\mathrm{hard} 
	+ \mathrm{sTr} \log\!  \big( X_{\mathbf{ab}}[\eta^s;\, G] + \mathcal{Q}_{\mathbf{ab}}^{H}[\overline{\phi},\, \overline{\phi};\, F_H] \big) \Big|_\mathrm{soft} \\
	- \mathrm{sTr} \log\! \big( X_{\mathbf{ab}}[\eta^s;\, G] + \mathcal{Q}_{\mathbf{ab}}^{H}[\overline{\phi},\, \overline{\phi};\, F_H'] \big) \Big|_\mathrm{soft}
	\Big].
	\end{multline}
With the factorization of the gauge-fixing for the heavy and light vectors enabled by the partial fixing, we can again choose $ F_H' = F_H $, such that the gauge-fixing terms of the unbroken group $ H $ coincide in the UV and EFT actions. The result is the recovery of the familiar matching formula 
	\begin{equation} \label{eq:part_fixed_matching_formula}
	S_\EFT^{(1)}[\overline{\phi};\, F_H,\, G] = \dfrac{i }{2} \mathrm{sTr} \log \mathcal{Q}^{\mathrm{all}}_{\mathbf{IJ}} [\eta^\mathrm{s},\, \overline{\phi};\, F_H,\, G] \Big|_\mathrm{hard}.
	\end{equation}
	
By explicit construction, we have, thus, been able to show that the partially fixed BF gauge enables the use of matching strategies based on identifying the one-loop EFT action with the hard-region UV loops in theories with spontaneously broken gauge symmetries. It allows for the same direct calculation of the gauge-invariant $ S_\EFT $ that is so convenient in the BF gauge for unbroken theories. 
The matching formula depends on the choices of both partial gauge-fixing condition and the BF gauge operator used to fix the unbroken $ H $. Nevertheless, all choices satisfying the outlined conditions may be used according to convenience. With the conditions suggested in Section~\ref{sec:part_fixed_gf_conditions}, calculations can be done with ordinary $ R_\xi  $-like propagators and but a few new interactions introduced by the gauge-fixing term.

\subsection{EFT matching with ordinary BF gauges}
Having seen how straightforward the adaptation of the matching formula is to the case of SSB using partially fixed BF gauges, we return to the problems associated with using an ordinary BF gauge for the UV theory. We examine the generic scenario of the last section where the massive vectors from the breaking $ G\to H $ have masses comparable to the heavy scale $ \Lambda $. The EFT BF effective action is given by Eq.~\eqref{eq:EFT_eff_action_H} in this case too.

\subsubsection{The UV effective action} 
We evaluate the effective action of the UV theory in the BF gauge, retaining explicit $ G $ invariance for the loop calculations. The effective action~\eqref{eq:one-loop_GI_effecitve_action} is 
	\begin{equation}
	\Gamma_\sscript{BF}[\overline{\eta}] = S[\overline{\eta}] + S^G_{\mathrm{bg.}}[\overline{\eta};\, \mathcal{F}_G] + \dfrac{i}{2} \mathrm{sTr} \log \mathcal{Q}^{\mathrm{all}}[\overline{\eta},\, \overline{\eta};\, F_G]
	\end{equation}
to one-loop order, where the fluctuation operator decomposes as 
	\begin{equation}
	\mathcal{Q}^{\mathrm{all}}_{\mathbf{IJ}}[\overline{\eta},\, \overline{\eta};\, F_G] = \mathcal{Q}_{\mathbf{IJ}}[\overline{\eta}] + \mathcal{Q}_{\mathbf{IJ}}^{G}[\overline{\eta},\, \overline{\eta};\, F_G],
	\end{equation} 
corresponding to $ S $ and $ S^G_\mathrm{fix} $, respectively.
In parallel to previous cases, the fluctuation operator can be written in block form as
	\begin{equation}
	\mathcal{Q}^{\mathrm{all}}_{\mathbf{IJ}}[\overline{\eta},\, \overline{\phi};\, F_G] = 	\begin{pmatrix}
		\mathcal{Q}^{\mathrm{all}}_{xy} & \mathcal{Q}^{\mathrm{all}}_{x b}\\
		\mathcal{Q}^{\mathrm{all}}_{a y} & \mathcal{Q}^{\mathrm{all}}_{ab}
	\end{pmatrix} \oplus
	\begin{pmatrix}
		\mathcal{Q}^{G}_{\textsf{xy}} & \mathcal{Q}^{G}_{\textsf{xb}} \\
		\mathcal{Q}^{G}_{\textsf{ay}} & \mathcal{Q}^{G}_{\textsf{ab}}
	\end{pmatrix}.
	\end{equation}
This time around, off-diagonal blocks are mixing the massive and massless ghosts, which may complicate the soft-region correspondence with the EFT ghost loops. There also exists an upper triangular matrix $ V\ud{\mathbf{I}}{\mathbf{J}} $ with determinant $ 1 $ that block-diagonalizes the fluctuation operator:
	\begin{equation} 
	X_{\mathbf{IJ}}
	\equiv V\ud{\mathbf{K}}{\mathbf{I}} \mathcal{Q}^{\mathrm{all}}_{\mathbf{KL}} V\ud{\mathbf{L}}{\mathbf{J}} = \begin{pmatrix}
		\mathcal{Q}^{\mathrm{all}}_{xy} & 0 \\ 0 & \mathcal{Q}^{\mathrm{all}}_{ab} - \mathcal{Q}^{\mathrm{all}}_{az} \mathcal{Q}^{\mathrm{all} \,\eminus 1}_{zu} \mathcal{Q}^{\mathrm{all}}_{ub} 
	\end{pmatrix} \oplus
	\begin{pmatrix}
		\mathcal{Q}^{G}_{\textsf{xy}} & 0 \\
		0 & \mathcal{Q}^{G}_{\textsf{ab}} -\mathcal{Q}^{G}_{\textsf{az}} \mathcal{Q}^{G\,\eminus 1}_{\textsf{uz}} \mathcal{Q}^{G}_{\textsf{zb}}
	\end{pmatrix}.
	\end{equation}

\subsubsection{One-loop matching}
For the last time, we apply matching condition~\eqref{eq:matching_condition_BF} at tree level to determine the leading EFT action: 
	\begin{equation}
	S_\EFT^{(0)}[\overline{\phi}] = S[\eta^\mathrm{s}] + S_\mathrm{bg.}^G[\eta^\mathrm{s};\, \mathcal{F}_G]- S_\mathrm{bg.}^H[\overline{\phi};\, \mathcal{F}_H].
	\end{equation}
In contrast to the previous approach, it is not possible to choose identical gauge-fixing conditions for the background fields, such that the terms drop; the two terms fix the $ G $ and the $ H $ groups, respectively. Despite this, we can still choose $ \mathcal{F}_G $ in a manner such that the resulting EFT action is invariant under $ H $. We merely let
	\begin{equation}
	S_\mathrm{bg.}^G[\overline{\eta};\, \mathcal{F}_G] = S_\mathrm{bg.}^{G/H}[\overline{\eta};\, \mathcal{F}_{G/H}] + S_\mathrm{bg.}^H[\overline{\phi};\, \mathcal{F}_H],
	\end{equation}
with the condition that the first term, gauge-fixing the massive vectors, is covariant under $ H $.\footnote{An example of such a background gauge-fixing is $ S_\mathrm{bg.}^{G/H}[\overline{\eta};\, \mathcal{F}_{G/H}] = \tfrac{1}{2\zeta} (\overline{d}_\mu \overline{V}_i^\mu - \zeta M_i \overline{\chi}_i) (\overline{d}^\mu \overline{V}_\mu^i - \zeta M_i \overline{\chi}_i) $.} Without this condition, we would not recover a gauge-invariant EFT action. The ordinary BF gauge is, thus, suitable for tree-level matching.

According to the matching condition~\eqref{eq:matching_condition_BF}, we are interested in the effective action, where the heavy fields are solutions to their EOMs in the presence of light background fields. At tree level, we have 
	\begin{equation}
	0 = \dfrac{\delta S'}{\delta \Phi^x}[\eta^\mathrm{s};\, \mathcal{F}_{G/H}], \qquad  S'[\overline{\eta};\, \mathcal{F}_{G/H}] = S[\overline{\eta}] + S_{\mathrm{bg.}}^{G/H}[\overline{\eta};\, \mathcal{F}_{G/H}].
	\end{equation}
This is our first indication that something may not work smoothly when using the ordinary BF gauge at loop level: this time the heavy field EOMs depend on the gauge-fixing condition for the background fields $ \mathcal{F}_{G/H} $ whereas their propagation in UV loops is determined by the quantum gauge-fixing condition $ F_G $.

The tree-level matching condition allows for the determination of the EFT fluctuation operator:
	\begin{equation}
	\mathcal{Q}_{\EFT \,ab}[\overline{\phi};\, \mathcal{F}_{G/H}] = \mathcal{Q}'_{ab} - \mathcal{Q}'_{ax} \mathcal{Q}^{\prime \eminus 1}_{xy} \mathcal{Q}'_{yb},
	\end{equation}
where, following the usual convention, $ \mathcal{Q}'[\eta^\mathrm{s};\, \mathcal{F}_{G/H}] $ is the fluctuation operator associated to $ S'[\eta;\, \mathcal{F}_{G/H}] $. Complete cancellation of soft-region EFT and UV loops in the matching condition, in the manner of Eq.~\eqref{eq:part_fixed_soft_cancellation}, now requires that
\begin{multline} \label{eq:bf_gauge_soft_cancellation}
	\mathrm{sTr} \log\!  \big( \mathcal{Q}^{\mathrm{all}}_{ab} - \mathcal{Q}^{\mathrm{all}}_{az} \mathcal{Q}^{\mathrm{all} \,\eminus 1}_{zu} \mathcal{Q}^{\mathrm{all}}_{ub} \big) \Big|_\mathrm{soft} +
	\mathrm{sTr} \log\!  \big( \mathcal{Q}^{G}_{\textsf{ab}} -\mathcal{Q}^{G}_{\textsf{az}} \mathcal{Q}^{G\,\eminus 1}_{\textsf{uz}} \mathcal{Q}^{G}_{\textsf{zb}}  \big) \Big|_\mathrm{soft} =\\
	\mathrm{sTr} \log\! \big( \mathcal{Q}'_{ab} - \mathcal{Q}'_{ax} \mathcal{Q}^{\prime \eminus 1}_{xy} \mathcal{Q}'_{yb} + \mathcal{Q}^{H}_{ab} \big) \Big|_\mathrm{soft} +
	\mathrm{sTr} \log \mathcal{Q}^{H}_\textsf{ab} \Big|_\mathrm{soft}.
\end{multline}
Can the various gauge-fixing conditions be chosen such as to ensure that this equality holds? We have not been able to make significant progress towards showing this. One might expect that the UV ghost loops should cancel against EFT ghost loops (going by the notion of a one-to-one correspondence). Then we would clearly have to choose $ F_H $ in a very particular manner to reproduce the effects of the integrated-out heavy ghosts in the UV loops. 
Such a choice will not prove a practical issue if it is possible to establish the matching condition, as in that event we would not have to compute EFT loops no matter how complicated the gauge might be.
On the other hand, cancellation of the trace over ordinary fields would seem to require that $ \mathcal{F}_{G/H} $ should be chosen in a manner compatible with $ F_G $ and $ F_H $ such that the heavy vectors are integrated out with similar propagators, as dictated by $ \mathcal{Q}^{\prime\, \eminus 1}_{xy} $ and $ \mathcal{Q}^{\mathrm{all}\, \eminus 1}_{xy} $, respectively. Perhaps then, one is not free to choose $ \mathcal{F}_{G/H} $ independently of $ F_G $. Thus, we find ourselves without anything more concrete than speculations about whether the soft-region cancellation~\eqref{eq:bf_gauge_soft_cancellation} can be attained in ordinary BF gauges.

\section{Conclusion and Outlook}
There is an emergent consensus that EFT matching calculations ought to be performed in the BF gauge with its gauge-invariant effective action. 
Nevertheless, we have found a conspicuous lack of justification for the use of the hard-region matching formula in gauge theories despite the maturity of the subject and the drastic increase in the number of one-loop matching calculations following the introduction of new tools. Hence, we found it timely to reexamine the subject and rectify this shortcoming.

For unbroken gauge theories, we were able to show the validity of the standard hard-region matching formula using BF gauges---just as expected. However, to our surprise, we found ourselves unable to demonstrate the cancellation of soft-region UV loops and EFT loops in spontaneously broken gauge symmetries using ordinary BF gauges. With the benefit of hindsight, it is clear that there would have to be some complications in equating the effective actions of the UV theory with the EFT in this case, seeing that the gauge group is larger for the UV theory. 
To our knowledge, a version of this setup has been used only once for practical calculations (including a proper gauge-fixing of the background fields), namely for the impressive matching of SMEFT to LEFT at one-loop order~\cite{Dekens:2019ept}. As a cross-check the authors verified that the result was independent of the gauge parameter in the on-shell basis, suggesting that the approach produced the correct result. Whether the validity of the approach generalizes beyond dimension-6 matching or to more complicated symmetry-breaking patterns remains unclear. The lack of a proof of general validity, suggests that further scrutiny is required before the use of ordinary BF gauges sees wider adoption in the matching of spontaneously broken gauge theories. 
It may seem like a disappointing conclusion to the paper that we were unable to prove the matching formula for this arrangement. Then again, why should we want to use such a complicated gauge? 
The partially fixed BF gauges offer much the same benefits as those promised by the ordinary BF gauges but with a significantly simplified gauge-fixing action to boot.

The partially fixed BF gauge is formulated here as an alternative to the ordinary BF gauge, maintaining explicit invariance of the effective action under background gauge variations w.r.t. the unbroken remnant group. 
Crucially, this gauge exhibits the correspondence between EFT and UV loops required to establish the hard-region matching formula at one-loop order~\cite{Fuentes-Martin:2016uol,Zhang:2016pja}. 
Nor do we foresee additional obstacles to extending the matching proof in the partially fixed BF gauges beyond one-loop order in the manner of Ref.~\cite{Fuentes-Martin:2023ljp}.
On a practical level, the partially fixed BF gauge generically provides simpler gauge-fixing terms compared to the ordinary BF gauge, making it significantly easier to set up calculations: 
there is no need to distinguish between background and quantum fields for the $ G/H $ fixing terms; the massless ghosts couple exclusively to massless vectors; and the fixing terms involve no quartic scalar interactions nor interactions between four massive vectors.
The gauge would, therefore, seem preferable to the ordinary BF gauge even if the matching formula could be established with the same level of rigor for the latter.

The results presented in this paper point the way toward EFT matching of broken gauge theories, which stands as the next frontier in the matching space. 
The features of the partially fixed BF gauge also make it ideal for implementation in automated matching codes. Nevertheless, there are still open questions about the role of gauge-fixing in matching calculations: not least, if it is possible to derive a hard-region matching formula for ordinary BF gauges. $ R_\xi $-like gauges, as we have used here, are also known to give masses to ghosts and would-be Goldstone bosons proportional to the gauge parameters. One might therefore presume that they could be made arbitrarily light, in which case they would not decouple along with the massive vectors. We would assume that the decoupling should be done before taking the limit $ \xi \to 0 $. Nevertheless, it would be interesting topic for further investigation.
This paper has been exploring matching under the assumption that massive vectors get masses comparable to the matching scale and decouple from the theory. It remains an open question of what to do if, e.g., the gauge couplings are small, or some Higgs field very heavy, such that the massive vectors remain dynamical in the EFT. Ref.~\cite{Dittmaier:1995cr,Dittmaier:2021fls} examined how to handle with this situation in electroweak symmetry breaking with a heavy Higgs boson, but work is still needed to examine how this approach generalizes to other symmetry-breaking scenarios.

\subsection*{Acknowledgments}
I am grateful to Javier Fuentes-Martín, Julie Pagès, Peter Stoffer, and Felix Wilsch for their helpful suggestions on how to improve the manuscript. I also want to thank Javier Fuentes-Martín, Jose Santiago, and Peter Stoffer for helpful discussions during the course of this work. This work was funded by the Swiss National Science Foundation (SNSF) through the Ambizione grant ``Matching and Running: Improved Precision in the Hunt for New Physics,'' project number 209042.

\renewcommand{\thesection}{\Alph{section}}
\appendix

\section{Extension to Curved Metrics} \label{app:curved_metric}
The discussion presented in this paper could have been generalized further, but we are sure the reader will forgive us for postponing a last generalization to the appendix. Until this point we have been assuming that the metric of the kinetic terms has been field-independent; however, this need not be the case for EFTs. Recently there has been much interest in formulating a geometric framework for EFTs, featured prominently in the GeoSMEFT program~\cite{Helset:2020yio}. It was suggested in Ref.~\cite{Helset:2018fgq} that one should account for the presence of higher-order terms in the SMEFT action to ensure that the gauge-invariant effective action of the BF gauge is also reparametrization invariant, as required by that framework. A similar idea can be adopted in the generic discussion of gauge-fixing discussed here.

The first place to include the field-dependence from higher order terms is for the metric $ a^{\eminus 1}_{AB} $ on the Lie algebra, which is promoted to $ a^{\eminus 1}_{AB}[\eta] $. It must respect gauge invariance, which translates to the requirement that 
	\begin{equation}
	0 = a^{\eminus 1}_{DB}[\eta] f\ud{D}{CA} \alpha^C + a^{\eminus 1}_{AD}[\eta] f\ud{D}{CB} \alpha^C + a^{\eminus 1}_{AB,I}[\eta] \big( D\ud{I}{C}[\eta]\alpha^C \big), \qquad \alpha \in \mathfrak{g}.
	\end{equation}
The gauge kinetic term is then 
	\begin{equation}
	S \supset - \dfrac{1}{4} A_{\mu\nu}^{A} a^{\eminus 1}_{AB}[\eta] A^{B \mu\nu}.
	\end{equation}
The scalar kinetic term is similarly equipped with a field-dependent metric $ h_{ab}[\eta] $ (replacing $ h_{ab} $) and reads 
	\begin{equation}
	S \supset (D_\mu \varphi^a) h_{ab}[\eta] (D^\mu \varphi^b).
	\end{equation}
To preserve gauge invariance of the action, the metric should be covariant, satisfying
	\begin{equation}
	0 = -i t^c_{Aa} \alpha^A h_{cb}[\eta] -i t^c_{Ab} \alpha^A h_{ac}[\eta] + h_{ab,I}[\eta] \big( D\ud{I}{A}[\eta]\alpha^A \big), \qquad \alpha \in \mathfrak{g}.
	\end{equation}
Including these quantities consistently should allow for the formulation of (partially fixed) BF gauges in a geometric framework.

\section{Faddeev--Popov Procedure} \label{app:FP_gauge_fixing} 

We include a brief review of the Faddeev--Popov approach to removing the gauge redundancy from the path integral for easy reference. Here we stick with the generic notation used elsewhere in this paper, but more exhaustive discussions couched in a more familiar setup can be found in most QFT textbooks. Our starting point is the partition function for a gauge theory with gauge group $ G $:
	\begin{equation} \label{eq:generic_partition_function}
	Z = \int \mathcal{D} \eta\, e^{iS[\eta]}, 
	\end{equation} 
having left out source terms for gauge-invariant operators. The gauge redundancy makes it impossible to invert the kinetic operator of the gauge field and there is no conjugate momentum to $ A^{A}_{\mu=0} $, meaning that it is not a dynamical variable. 

To proceed, one introduces a gauge-fixing condition in the form of a function $ G[\eta] \in \mathfrak{g} $, such that 
	\begin{equation} \label{eq:generic_gauge_cond}
	G[\eta_g] = 0 \qquad \text{for a unique $ g=g_0[\eta] $}  
	\end{equation}
for all possible $ \eta $. The notation $ \eta_g $ is used to denote the transformation by a group element $ g(x) $. 
In the vicinity $ g\sim g_0 $ of the unique solution to Eq.~\eqref{eq:generic_gauge_cond}, we may parametrize the group elements by $ g= (1 + \alpha) g_0 $ for $ \alpha\in \mathfrak{g} $. Thus, $ \eta_g^{I} = \eta_{g_0}^{I} + D\ud{I}{A}[\eta_{g_0}] \alpha^A $. This allows for evaluating the integral over the delta function of the gauge-fixing condition as 
	\begin{equation} \label{eq:int_delta_gf}
	\int_{G} \dd \mu[g]\, \delta\big(G[\eta_g]\big) = \int \mathcal{D} \alpha\,  \delta\big(G\ud{A}{\!,I}[\eta_{g_0}] D\ud{I}{B}[\eta_{g_0}] \alpha^B \big) = \mathrm{Det}^{\eminus 1} \big( G\ud{A}{\!,I}[ \eta_{g_0}] D\ud{I}{B}[ \eta_{g_0}] \big)
	\end{equation}
given some $ \eta $. Here $ \mu[g] $ denotes the Haar measure. 

We can recast the partition function~\eqref{eq:generic_partition_function} as 
	\begin{equation} \label{eq:generic_gf_partition_function}
	\begin{split}
	Z &= \int \mathcal{D} \eta \, \mathrm{Det} \big( G\ud{A}{\!,I}[ \eta_{g_0}] D\ud{I}{B}[ \eta_{g_0}] \big) \! \int_{G} \dd \mu[g]\, \delta\big(G[\eta_g]\big) \, e^{iS[\eta]} \\
	&= \int_{G} \dd \mu[g] \! \int \mathcal{D} \eta \, \delta\big(G[\eta]\big) \, \mathrm{Det} \big( G\ud{A}{\!,I}[ \eta] D\ud{I}{B}[ \eta] \big) \, e^{iS[\eta]}.
	\end{split}
	\end{equation}
We utilized that the delta function picks out the group element such that $ \eta_{g_0} = \eta_g $ while the measure is invariant under $ \mathcal{D} \eta = \mathcal{D} \eta_{g} $. It also holds that $ S[\eta_g] = S[\eta] $ from the invariance of the action. Following standard arguments, the functional determinant is cast as a path integral over ghost fields:
	\begin{equation}
	\mathrm{Det} \big( G\ud{A}{\!,I}[ \eta] D\ud{I}{B}[ \eta] \big) 
	=\int \mathcal{D} \boldsymbol{\omega}\, \exp\left[ -i\! \int_x \overline{\omega}_A G\ud{A}{\!,I}[A] D\ud{I}{B}[\eta]\, \omega^B \right].
	\end{equation}
Accordingly, we arrive at the partition function 
	\begin{equation} \label{eq:delta_function_path_integral}
	Z = \int \mathcal{D} \eta \, \mathcal{D} \boldsymbol{\omega}\, \delta\big(G[\eta]\big)  \exp \left[i \left(S[\eta] - \! \int_x \overline{\omega}_A\, G\ud{A}{\!,I}[A] D\ud{I}{B}[\eta]\, \omega^B \right) \right],
	\end{equation}
by dropping the volume factor from the integration over $ G $. The drawback of this direct gauge-fixing with a delta function in the path integral is that it is \emph{non-renormalizable}~\cite{Ferrari:2013aza}. This is usually avoided by shifting $ G[\eta] \to G[\eta] - \theta $ for some variable $ \theta\in\mathfrak{g} $. The partition function is independent of $ \theta $ or for that matter the choice of gauge-fixing condition. A final Gaussian integration $ \int\, \mathcal{D} \theta \, e^{\eminus \tfrac{i}{2\xi} \int_x\, \langle\theta,\, \theta \rangle} $ leaves the partition function
	\begin{equation} \label{eq:Gaussian_weight_gf_z}
	Z = \int \mathcal{D} \eta \, \mathcal{D} \boldsymbol{\omega}\, \exp \left[i \left(S[\eta] - \int_x \overline{\omega}_A G\ud{A}{\!,I}[A] D\ud{I}{B}[\eta]\, \omega^B - \frac{1}{2\xi} \int_x \brakets{G[\eta],\, G[\eta]} \right) \right],
	\end{equation}
ignoring an irrelevant overall normalization. The last two terms in the exponential constitute the gauge fixing terms for the action.

\section{Mass Basis of a Spontaneously Broken Gauge Theory} \label{app:broken_phase}
Here we give a few additional details that are helpful for going to the mass basis in a broken gauge theory. The expectation is that this will prove useful for the practitioner wishing to implement the partially fixed BF gauge in a concrete model.

\subsection{Yang--Mills action}
The algebra of the gauge group in the broken phase has been discussed in, e.g., Ref.~\cite{Henning:2014wua}, and we adapt their discussion here.  
In the mass basis, the structure constant reads
	\begin{equation}
	0 = f\du{\alpha}{a} = -i\, t_{\alpha b}^{a} v^b,\qquad  0 \neq f\du{i}{a} = -i\, x_{ib}^{a} v^b,
	\end{equation}
where $ f\du{i}{a} $ are linearly independent for different $ i $ (otherwise some of the heavy vectors would, in fact, be massless). By multiplying $ f\du{\alpha}{a} $ with an unbroken generator $ t_\beta $, it becomes clear that the algebra of the unbroken generators, $ t_\alpha $, closes:
	\begin{equation}
	\commutator{t_{\alpha}}{t_{\beta}} = i\, t_{\gamma}  f\ud{\gamma}{\alpha\beta}.
	\end{equation}
We identify it with the Lie algebra $ \mathfrak{h} $ of the unbroken group $ H\subseteq G $.

From the cyclic property of the metric on $ \mathfrak{g} $ and closure of $ \mathfrak{h} $, it follows that 
	\begin{equation}
	\brakets{t_{\beta},\, \commutator{t_{\alpha}}{x_i}} =  \brakets{x_i,\, \commutator{t_\beta}{t_\alpha}} = 0,
	\end{equation}
which in turn indicates that 
	\begin{equation}
	\commutator{t_\alpha}{x_i} = i x_j f\ud{j}{\alpha i}.
	\end{equation}
Thus, we can verify that the heavy vectors transform in some faithful representation of the unbroken group $ H $ under the group action:
	\begin{equation} \label{eq:heavy_V_representation}
	\delta_\alpha^H V_\mu = i \alpha^\alpha \commutator{t_\alpha}{V_\mu} = - x_i \alpha^\alpha f\ud{i}{\alpha j} V^j_\mu.
	\end{equation}
The covariant derivative decomposes as in Eq.~\eqref{eq:covD_heavy-light_split}. This implies, that the covariant derivative of the unbroken group acts on $ V_\mu $ as
	\begin{equation}
	d_\mu V_\nu = \commutator{d_\mu}{V_\nu} = x_i (\partial_\mu V_\nu^i + f\ud{i}{\alpha j} B_\mu^\alpha V_\nu^j).  
	\end{equation}
The field-strength tensor decomposes as
	\begin{equation}
	A_{\mu\nu} = i \commutator{d_\mu - i V_\mu}{d_\nu - i V_\nu} = B_{\mu\nu} + V_{\mu\nu} - i \commutator{V_\mu}{V_\nu},
	\end{equation}
where 
	\begin{equation}
	B_{\mu\nu} = i[d_\mu,\, d_\nu], \qquad V_{\mu\nu} = d_{\mu} V_\nu - d_{\nu} V_\mu
	\end{equation}
are the field strength tensors of the $ B_\mu $ and $ V_\mu $ fields, respectively. With this notation, the Yang-Mills action reads
	\begin{align}
	\L_\sscript{YM}
	&= -\dfrac{1}{4} \brakets{B_{\mu\nu},\, B^{\mu\nu}} - \dfrac{1}{4}V^{i}_{\mu\nu} V^{\mu\nu}_i - \dfrac{1}{2} \big(  B^{\alpha\, \mu\nu} \hat{a}_{\alpha \beta}^{\eminus 1} f\ud{\beta}{ij}  +  V_k^{\mu\nu} f\ud{k}{ij} \big) V_\mu^{i} V_\nu^{j} \nonumber\\
	&\qquad  -\dfrac{1}{4} V_\mu^{i} V_\nu^{j} V^{\mu k} V^{\nu \ell} \big(f\ud{\alpha}{ij} \hat{a}_{\alpha \beta}^{\eminus 1} f\ud{\beta}{k\ell} + f\ud{m}{ij} f_{mk\ell}\big)
	\end{align}
in the broken phase.

As mentioned some care has to be taken in the case where some of the massive vectors arrange themselves into complex fields. One could deal with this issue in the manner of the complex scalars and separate the real and imaginary components, however, a more compelling option emerges. We may count both the field and its complex conjugate among the degrees of freedom at the same level. Let us consider the familiar example of electroweak symmetry breaking: We arrange the massive vectors as $ V^i_\mu \supset \big( Z_\mu,\, W_\mu^+,\, W_\mu^- \big) $.
In this example, the mass matrix can be written as 
	\begin{equation}
	\kappa_{ij} M_j^2 = 
	\begin{pmatrix} m_Z^2 & & \\ & & m_W^2 \\ & m_W^2 &
	\end{pmatrix}, \qquad  \kappa_{ij}  = 
	\begin{pmatrix} 1 & & \\ & & 1 \\ & 1 &
	\end{pmatrix},
	\end{equation}
and we merely have to include the metric $ \kappa $ as we have done in this paper to accommodate the complex $ W^{\pm}_\mu $ fields. The masses are always such that 
	\begin{equation}
	M_i \kappa_{ij} = \kappa_{ij} M_j,
	\end{equation}
as the off-diagonal masses in the mass matrix come from pairs of complex fields.

\subsection{Goldstone bosons and scalar sector} \label{sec:GBs}
The would-be Goldstone bosons---henceforth referred to as GBs for brevity---of the broken theory correspond to the flat directions around the minimum of the potential. We can, therefore, embed them into the scalar multiplet along the independent $ x_{ib}^{a} v^b $ directions. We decompose\footnote{Recall that the scalar metric $ h_{ab} $ is used to lower indices where appropriate.}
	\begin{equation} \label{eq:GB_embedding1}
	\varphi^a = \varphi^a_h + \varphi_\chi^a, \qquad \varphi_{h\,a} f\du{i}{a} = 0 \quad \forall i,  
	\end{equation} 
choosing the Higgs fields to be orthogonal to the Goldstone directions. We parametrize the GBs by 
	\begin{equation} \label{eq:GB_embedding2}
	\varphi^a_\chi = - i \tilde{\chi}^i x_{ib}^{a} v^b = \tilde{\chi}^i f\du{i}{a} 
	\end{equation}
and, thus, $ \varphi_{h\, a} \varphi_\chi^a  = 0 $. The transformation of the GBs under the remnant group $ H $ follows from 
	\begin{equation}
	t_{\alpha b}^{a} \varphi_\chi^b = -i \tilde{\chi}^i \commutator{t_\alpha}{x_i}\ud{a}{b} v^b = \tilde{\chi}^j f\ud{i}{\alpha j} x_{i b}^{a} v^b = (i f\ud{i}{\alpha j} \tilde{\chi}^j) f\du{i}{a},
	\end{equation}
which demonstrates that $ \tilde{\chi}^i $ transforms in the same representation as $ V_\mu^i $ in Eq.~\eqref{eq:heavy_V_representation}. It is easy to demonstrate that $ t_{\alpha b}^{a} \varphi^h_b $ is orthogonal to $ \varphi^a_\chi $, and we conclude that also $ \varphi^h_a $ constitute a representation of $ H $.

Recall that the mass matrix~\eqref{eq:heavy_V_mass_matrix} of the heavy fields is given by 
	\begin{equation}
	M^2_{i} \kappa_{ij} = M_{ij}^{2}= f_{ia} f\du{j}{a}.
	\end{equation} 
This leads us to identify the canonically normalized GBs as 
	\begin{equation}
	\chi^i = \tilde{\chi}^i M_i\implies \varphi_a^\chi \varphi_a^\chi = \chi_i \chi^i. 
	\end{equation}
We observe that the decay constant satisfies
	\begin{equation}
	f_{ia} \varphi^a = f_{ia} \varphi_\chi^a = M_{i} \chi_i.
	\end{equation}
It also holds that
	\begin{equation}
	\varphi^a (h f\transpose a f h)_{ab} \varphi^b = M_i^2 \chi_i \chi^i,
	\end{equation}
which will come in handy for the gauge-fixing terms. 

The scalar kinetic term
	\begin{equation}
	\mathcal{L}_\mathrm{kin} = \tfrac{1}{2} D_\mu \varphi^{\prime}_a D_\mu \varphi^{\prime a}
	\end{equation}
is gauge invariant, which is ensured by the covariance of scalar metric $ h_{ab} $, which satisfy
	\begin{equation}
	0= T^{c}_{Aa} h_{cb} +h_{ac} T^{c}_{Ab}.
	\end{equation}	
In the broken phase, the $ G $-covariant derivative of the scalar field decomposes as 
	\begin{equation}
	D_\mu \varphi^{\prime a}= d_\mu \varphi^a - i V^i_\mu x_{ib}^{a} \varphi^b  + V_\mu^i f\du{i}{a}.
	\end{equation}
Thus, the scalar kinetic term reads
	\begin{equation} \label{eq:scal_kin_term}
	\begin{split}
	\mathcal{L}_\mathrm{kin} 
	&= \tfrac{1}{2} d_\mu \varphi^a_h h_{ab} d^\mu \varphi^b_h + \tfrac{1}{2} d_\mu \chi_i d^\mu \chi^i + \tfrac{1}{2} M^2_{i} V_i^{\mu} V^i_\mu + d^\mu\chi_i M_{i} V^i_\mu \\
	&\quad + \tfrac{1}{2} i V_\mu^i (\varphi_a  x^{a}_{ib} \overset{\leftrightarrow}{d^\mu} \varphi^b) + \tfrac{1}{2} V_\mu^i V^{j\mu} \varphi_a (x_i x_j \varphi)^a -i V_\mu^i V^{j\mu} f_{ia}  x^{a}_{jb} \varphi^b,
	\end{split}
	\end{equation}
where $ \varphi_a \overset{\leftrightarrow}{d^\mu} \varphi^b = \varphi_a (d_\mu \varphi^b) - (d_\mu \varphi_a) \varphi^b $. All quadratic terms involving scalar fields neatly separate into Higgs and GB components, allowing for the identification of kinetic terms for the scalar mass eigenstates, the mass terms for the massive vector fields, and the kinetic mixing of those vectors with the GBs. On the other hand, interactions in the second line will have to be expanded out in each specific model, as no obvious general simplifications are possible.

\section{Broken Phase of a $ \SU(3) \to \SU(2) $ Toy Model}
\label{app:toy_model}
This appendix spells out the broken phase of a toy model to elucidate our notation and showcase the formalism for practitioners. In a compromise between complexity and triviality, we have chosen to focus on the breaking $ G= \SU(3) \to H= \SU(2) $ by a scalar $ \Sigma' $ in the fundamental (triplet) representation of $ G $. In contrast to the breaking in the more familiar electroweak SM, this example includes a non-Abelian remnant group, which is found in many BSM models and introduces added complexity. We can also demonstrate the slightly nebulous inclusion of complex DOFs in our real multiplets through the use of non-identity metrics. This might seem a slight overkill for the scalars in this example given that there is only a single complex field in this example; the benefits come from being able to simultaneously treat mixed real and complex fields. 

The Lagrangian of the toy model is given by 
	\begin{equation}
	\L = -\dfrac{1}{4 g^2} A^{A}_{\mu\nu} A^{A\mu\nu} + D_\mu \Sigma^{\prime \dagger} D^\mu \Sigma^{\prime} - V( \Sigma^{\prime \dagger} \Sigma'),
	\end{equation}
where $ V $ is the scalar potential. To make connection with the notation introduced in Sec.~\ref{sec:broken_gauge_theory}, we may identify 
	\begin{equation}
	a^{\eminus 1}_{AB} = g^{\eminus 2} \delta_{AB}, \qquad  \varphi^{\prime a} = (\Sigma',\, \Sigma^{'\ast}) ,\qquad h_{ab} = \begin{pmatrix}
	& \mathds{1} \\ \mathds{1} &
	\end{pmatrix} ,
	\end{equation}
where the metric on the Lie algebra is proportional to the identity matrix, since $ G $ is a simple group, and $ g $ is the gauge coupling constant associated with $ G $. The specifics of the scalar potential are largely irrelevant, but we assume that it is such that $ \Sigma'  $ develops a VEV $ \langle \Sigma' \rangle = (0, 0, v) $, which acts as an order parameter breaking $ G $ to $ H $.

\subsection{Algebra of the broken group}
The adjoint representation of $ G $ decomposes as $ \rep{8} \to \rep{3} \oplus \rep{2} \oplus \repbar{2} \oplus \rep{1} $ under the unbroken group, where we use $ \repbar{2} $ to denote the conjugate doublet representation for bookkeeping purposes even though it is isomorphic to the fundamental representation $ \rep{2} $. In terms of the gauge fields $ A^A_\mu $, the $ \rep{3} $ representation will correspond to the massless field $ B^\alpha_\mu = (W^\alpha_\mu) $ of $ H $. The remaining representations are associated with heavy vectors, which, in a slight abuse of index notation, read $ V^i_\mu = (X^i_\mu,\, X^\ast_{i\mu},\, Z_\mu) $; that is, the two doublets form a single complex field. Since we are working from a simple group $ G $, the gauge coupling of the remnant group is identified with that of $ G $, so we can immediately determine that the metrics of the broken phase are given by (the form of $ \kappa_{ij} $ follows from the prescription)
	\begin{equation}
	\hat{a}^{\eminus 1}_{\alpha \beta}= g^{\eminus 2} \delta_{\alpha\beta}, \qquad \kappa_{ij} = \begin{pmatrix}
	0 & \delta\ud{i}{j} & \\ \delta\du{i}{j} & 0& \\ & & 1
	\end{pmatrix}.
	\end{equation}

If we turn to the splitting of generators into those associated with $ B^\alpha_\mu $ and $ V^i_\mu $, the present model is so simple that there is no mixing between multiple instances of the same irreducible representations of $ H $. Thus, we may simply guess their form from the requirement of $ H $-invariance without any need to construct the transformation matrix $ L\ud{A}{B} $:
The fundamental representation of $ G $ decomposes according to $ \rep{3} \to \rep{2} \oplus \rep{1} $ into a doublet and a singlet of $ H $. Hence, we can decompose the generators of the fundamental representation of $ G $ according to
	\begin{equation}
	T_A \to \begin{pmatrix}
	\tau^p_{\alpha q} & \\ & 0
	\end{pmatrix} \oplus 
	\dfrac{1}{\sqrt{2}} \begin{pmatrix}
	0 & \delta\ud{p}{i} \\ & 0
	\end{pmatrix} \oplus 
	\dfrac{1}{\sqrt{2}} \begin{pmatrix}
	0 &  \\ \delta\ud{i}{q} & 0
	\end{pmatrix} \oplus
	\dfrac{1}{\sqrt{12}} \begin{pmatrix}
	\delta\ud{p}{q} & \\ & -2
	\end{pmatrix}
	\end{equation}
where $ \tau_a = \tfrac{1}{2} \sigma_a $, $ \sigma_a $ being the Pauli matrices, and the indices $ p,q $ are doublet indices. The normalization of each term follows from normalizing $ \tr[T_A T_B] = \tfrac{1}{2} \delta_{AB} $. 
By `$ \oplus $' we have in mind the the decomposition of the adjoint index $ \phantom{}_{A} \to \phantom{}_{\alpha} \oplus \phantom{}_{i} \oplus \phantom{}^{i} \oplus \rep{1} $ in parallel with $ \rep{8} \to \rep{3} \oplus \rep{2} \oplus \repbar{2} \oplus \rep{1} $. The proper 3-index $ 8\times 3 \times 3 $ tensor can be assembled from the pieces.
Recalling the prescription that the gauge couplings are absorbed into the generators associated with $ V^i_\mu $, we identify
	\begin{equation} \label{eq:generator_decomposition}
	t_\alpha = \begin{pmatrix}
	\tau^p_{\alpha q} & \\ & 0
	\end{pmatrix},\qquad 
	x_i = \dfrac{g}{\sqrt{2}} \begin{pmatrix}
	0 & \delta\du{i}{p} \\ & 0
	\end{pmatrix} \oplus 
	\dfrac{g}{\sqrt{2}} \begin{pmatrix}
	0 &  \\ \delta\ud{i}{q} & 0
	\end{pmatrix} \oplus
	\dfrac{g}{\sqrt{12}} \begin{pmatrix}
	\delta\ud{p}{q} & \\ & \eminus 2
	\end{pmatrix}
	\end{equation}
Whereas the index $ \alpha $ of the unbroken generators takes values in $ \rep{3} $, the $ i $ index of the broken generators takes values in $ \repbar{2} \oplus \rep{2} \oplus \rep{1} $.
We will see later that this identification does, in fact, lead to diagonal masses as it should. It is worth pointing out, that $ L\ud{A}{B} $ by necessity involves a unitary but not orthogonal rotation to organize the doublets into a single complex field. For this reason $ x_i $ is not Hermitian even though $ T_A $ is. 
To be more explicit, it follows from the Eq.~\eqref{eq:generator_decomposition} that the gauge bosons contracted on the generators of the fundamental $ G $ representation read
	\begin{equation}
	T_A A^{A}_\mu = t_\alpha B^\alpha_\mu + x_i V^i_\mu = \begin{pmatrix}
	B^\alpha_\mu \tau^p_{\alpha q} + \tfrac{g}{\sqrt{12}} Z_\mu \delta\ud{p}{q} &
	\tfrac{g}{\sqrt{2}} X^{p}_\mu \\
	\tfrac{g}{\sqrt{2}} X^{\ast}_{q\mu} &
	\eminus \tfrac{g}{\sqrt{3}} Z_\mu 
	\end{pmatrix}
	\end{equation}
in the broken basis.

Without an explicit construction of  $ L\ud{A}{B} $, it may seem daunting to recover the decomposition of the structure constants of $ G $; however, we may use the commutation rules for the generators along with their deconstruction. The broken phase structure constants are determined through relations such as\footnote{The diagonalization of the gauge fields ensure that $ \tr[t_\alpha x_i] = 0 $.}
	\begin{equation}
	\tr[x_i x_j] f\ud{j}{\beta k} = -i \tr \big[[x_i,\, t_\beta] x_k \big],
	\end{equation}
which follow from applying $ L\ud{A}{B} $ to the commutation relation
	\begin{equation}
	\tr[T_A T_D] f\ud{D}{BC} = -i \tr \big[[T_A,\, T_B] T_C \big].
	\end{equation}
Thus, we find that  
	\begin{equation}
	f\ud{\alpha}{jk} = i \hat{a}^{\alpha \beta} \left( \begin{pmatrix}
	0 & \eminus \tau^k_{\beta j} & \\ \tau^j_{\beta k} & 0 & \\ & & & 0
	\end{pmatrix}\right),
	\end{equation}
which is understood as a $ 1\times 3\times 3 $ tensor in the representations included in the broken-phase indices $ \alpha,j,k $ of the structure constant; namely $ \alpha $ assumes values only in $ \rep{3} $, whereas $ j,k $ takes values in the three representations $ \repbar{2}, \rep{2}, \rep{1} $ spanned by the massive vectors.\footnote{Again, in the abuse of notation we use the same index labels on both sides of the equation. The collective indices on the left should be understood as indexing into one of the representations and then into a specific index of that representation. That is, $ j,k\in \{ \repbar{2}_{i}, \,\rep{2}^i, \, \rep{1} \} $ and $ \alpha \in \{\rep{3}_\alpha \} $.} 
Similarly, there are the $ 3\times 3\times 1 $ tensors 	
	\begin{equation}
	f\ud{i}{j\gamma} = - f\ud{i}{\gamma j} = i  \left( \begin{pmatrix}
	\tau_{\gamma j}^i \\ 0 \\ 0
	\end{pmatrix}, \begin{pmatrix}
	0 \\ \eminus \tau_{\gamma i}^j \\ 0
	\end{pmatrix} , \begin{pmatrix}
	0 \\ \phantom{,}0\phantom{,} \\ 0
	\end{pmatrix} \right).
	\end{equation}
Observe here that $ f\ud{\alpha}{jk} $ includes a $ g^2 $ from $ \hat{a}_{\alpha \beta} $ whereas, $ f\ud{i}{j\gamma} $ are purely numbers. Lastly, the structure constant with three indices from the broken algebra is the $ 3\times 3 \times 3 $ tensor
	\begin{equation}
	f\ud{i}{jk}  = \dfrac{i\sqrt{3}}{2} g   \left( \begin{pmatrix}
	0 & & \delta\ud{i}{j} \\ & 0 & \\ \eminus \delta\ud{i}{k} && 0
	\end{pmatrix}, \begin{pmatrix}
	0 && \\ &0 & \eminus \delta\ud{j}{i} \\ & \delta\ud{k}{i} & 0 
	\end{pmatrix} , \begin{pmatrix}
	0 & \eminus \delta\ud{k}{j}&  \\ \delta\ud{j}{k} & 0 & \\ && 0
	\end{pmatrix} \right).
	\end{equation}
We can now proceed to the scalar sector.

\subsection{Scalar sector}
The index $ a $ of $ \varphi^{\prime a} $ runs over a fundamental and an anti-fundamental representation of $ G $. In the broken phase, each of these decomposes into a doublet and a singlet representation, so the scalar metric becomes
	\begin{equation}
	h_{ab} = \begin{pmatrix}
	& & \delta\du{q}{p} & \\ & & & 1 \\ \delta\ud{q}{p} & & & \\ & 1 & & 
	\end{pmatrix},
	\end{equation}
where again $ p,q $ are doublet indices. The generators acting on the (reducible) representation of $ \varphi^{\prime a} $ then reads
	\begin{equation}
	t^a_{\alpha b} = \begin{pmatrix}
	\tau^p_{\alpha q} & &&\\ & 0 && \\ && \eminus\tau^q_{\alpha p} & \\ &&& 0
	\end{pmatrix}
	\end{equation}
and
	\begin{equation}
	x^a_{ib} = \dfrac{g}{\sqrt{2}} \begin{pmatrix}
	0 & \delta\du{i}{p} && \\ & 0 && \\&&0& \\ & & \eminus \delta\ud{q}{i} &0
	\end{pmatrix} \oplus 
	\dfrac{g}{\sqrt{2}} \begin{pmatrix}
	0 & & & \\ \delta\ud{i}{q} & 0 & & \\ & & 0& \eminus \delta\du{p}{i} \\ &&&0
	\end{pmatrix} \oplus
	\dfrac{g}{\sqrt{12}} \begin{pmatrix}
	\delta\ud{p}{q} & &&\\ & \eminus 2 && \\  && \eminus \delta\du{p}{q}&\\ && & 2
	\end{pmatrix}.
	\end{equation}
Also here have we used `$ \oplus $' to denote the decomposition of the broken part of the adjoint index $ i $ into the representations $ \rep{2} $, $ \repbar{2} $, and $ \rep{1} $ of $ H $.

The VEV of the scalar field reads 
	\begin{equation}
	v^a = \big(0,\, v,\, 0,\, v\big)
	\end{equation}
and between that and the broken generators, the decay constant matrix~\eqref{eq:decay_constant} reads
	\begin{equation}
	f\du{i}{a} = \dfrac{i\, v\, g}{\sqrt{6}} \begin{pmatrix}
	\eminus\sqrt{3} \delta\du{i}{p} & 0 & 0 & 0 \\
	0 & 0 & \sqrt{3}\delta\ud{i}{p} \\ 
	0 & \sqrt{2} & 0 & \eminus \sqrt{2}
	\end{pmatrix} ,\qquad f\du{\alpha}{a} =0. 
	\end{equation}
We can now verify that the mass matrix is properly diagonalized in that
	\begin{equation}
	f\du{i}{a} h_{ab} f\du{j}{b} = M_i^2 \kappa_{ij}, \qquad M_i = \dfrac{g\, v}{\sqrt{6}} \big(\sqrt{3},\, \sqrt{3},\, 2\big),
	\end{equation}
from which the masses of the massive vectors are determined. Next, we introduce the would-be GBs associated with $ V^i_\mu $, letting $ \chi^i = \big(\chi_X^i,\, \chi^\ast_{Xi},\, \chi_Z\big) $. The GB bosons along with the singlet ``Higss'' field are embedded into $ \varphi^a $ according to
	\begin{equation}
	\varphi_\chi^a = \dfrac{i}{\sqrt{2}} \big(\eminus \sqrt{2} \chi_X^i,\, \chi_Z,\, \sqrt{2} \chi^\ast_{Xi},\, \eminus \chi_Z	 \big),\qquad \varphi_h^a= \dfrac{1}{\sqrt{2
	}} \big(0,\, h,\, 0,\,  h\big),
	\end{equation}
as per Eqs.~(\ref{eq:GB_embedding1}, \ref{eq:GB_embedding2}). The embedding of the GBs along with orthogonality and canonical normalization uniquely fixes $ h $ in $ \varphi_h^a $.

\subsection{Partial gauge-fixing terms}
We now have all the building blocks needed to write down the gauge-fixing terms for the toy model. Here we refer the result only in the partial fixed BF gauge, as this is the simpler gauge.
The gauge-fixing Lagrangian for the massive vectors~\eqref{eq:part_fixed_vectors} is 
	\begin{multline} \label{eq:toy_model_vector}
	\L^{G/H}_{\mathrm{vec.}} = -\dfrac{1}{\zeta} d_\mu X^{\ast \mu}_i d^\nu X^i_\nu - \dfrac{1}{2\zeta} \big(\partial_\mu Z^\mu \big)^2 + M_X \big( \chi^\ast_{Xi} d^\mu X^i_\mu \hc \big) \\
	+ M_Z \chi_Z \partial^\mu Z_\mu - \zeta M_X^2 \chi^\ast_{Xi} \chi_X^i - \dfrac{\zeta}{2} M^2_Z  \chi_Z^2.
	\end{multline}
The partial fixing also introduces ghost terms for ghost fields associated with the massive vectors. We denote the ghosts by 
$ u^i = \big(u_X^i, \, u^\ast_{Xi},\, u_Z \big) $ and the anti-ghost fields by $ \overline{u}_i = \big(\overline{u}_{Xi}, \, \overline{u}^{\ast i}_{X},\, \overline{u}_Z \big) $. Expanding out Eq.~\eqref{eq:part_fixed_ghosts} then yields the ghost terms
	\begin{align} \label{eq:toy_model_ghost}
	\L^{G/H}_{\mathrm{gh.}}& =\, - \overline{u}_{Xi} (d^2 + \zeta M^2_X) u_X^i - \overline{u}^{\ast i}_{X} (d^2 + \zeta M^2_X) u^\ast_{Xi} - \overline{u}_{Z} (\partial^2 + \zeta M^2_Z) u_Z \nonumber \\
	& +g \dfrac{\sqrt{3}}{2} \Big(i Z_\mu \overline{u}_{Xi} d^\mu u^i_X - i X^i_\mu \overline{u}_Z d^\mu u^{\ast}_{Xi} + iX^\ast_{i\mu} \overline{u}_X^{\ast i} \partial^\mu u_Z  \hc \Big)  \nonumber \\
	& + \dfrac{g^2}{4} \left(X_{i\mu}^\ast X_{j}^{\ast\mu} \overline{u}_X^{\ast i} u_X^j + X_\mu^i X^{\ast \mu}_j \overline{u}_{Xi} u_X^j - X_\mu^i X^{\ast \mu}_i \overline{u}_{Xj} u_X^j \hc\right) \nonumber \\
	& -\zeta \dfrac{g^2 v}{2} \left( \dfrac{h - i\, \chi_Z}{\sqrt{2}} \overline{u}_{Xi} u_X^i - \dfrac{i}{\sqrt{6
	}}\chi_X^i \overline{u}_Z u^\ast_{Xi} - i \sqrt{\dfrac{2}{3} } \chi^\ast_{Xi} \overline{u}^{\ast i}_{X} u_Z \hc \right)  \nonumber \\
	&- \zeta \dfrac{\sqrt{2}  g^2 v}{3} h\,  \overline{u}_Z u_Z 
	+\zeta \dfrac{g^2}{4} \left(2 \overline{u}_X^{\ast i} u^\ast_{Xj} \, \overline{u}_{X i} u_X^j - \overline{u}_X^{\ast i} u^\ast_{Xi} \, \overline{u}_{X j} u_X^j \right) \nonumber \\
	&  
	-\zeta \dfrac{g^2}{8} \left(2 \overline{u}_{Xi} u_X^j \, \overline{u}_{Xj} u_X^i - \overline{u}_{Xi} u_X^i \, \overline{u}_{Xj} u_X^j \hc \right).
	\end{align}
The quartic ghost terms of the last two lines can be safely ignored for the purposes of one-loop calculations. Neither the massive ghost nor vector gauge-fixing terms discriminate between background and quantum fields, drastically reducing the number of terms. 
The gauge-fixing terms for the massless vectors and ghost fields of the remnant group $ H $ are entirely standard and can be read off from Eqs.~(\ref{eq:BF_vectors}, \ref{eq:BF_ghosts}) using the appropriate group factors.

\section{The BF $ R_\xi $-Gauge Lagrangian in the Broken Phase} \label{app:broken_phase_BF-gauge}
We have referenced the complexity of using BF gauges in theories with SSB several times in this paper. To illustrate this point, we will refer the usual \emph{background field $ R_\xi $-gauge}~\cite{Shore:1981mj} in the broken phase of a generic theory in this appendix. Comparing the resulting formulas with the gauge-fixing terms of Section~\ref{sec:part_fixed_gf_conditions} easily demonstrates the point. This gauge has been employed for the SM in, e.g., Refs.~\cite{Einhorn:1988tc,Denner:1994xt,Helset:2018fgq}; however, we have not found the generic expression referred anywhere. The reader might find them useful for other purposes too.

\subsection{Gauge-fixing condition for the BF $ R_\xi $-gauge}
The gauge group is spontaneously broken by the VEVs of scalar fields $ \varphi^a $ as described in Section~\ref{sec:broken_gauge_theory}. The background fields are introduced by the shift $ \eta^I \to \eta^I + \overline{\eta}^I $ as per usual, with the one subtlety that the VEV of the scalar fields is put in the background fields:
	\begin{equation}
	\varphi^{\prime a} \longrightarrow \varphi^a + \overline{\varphi}^{\prime a}, \qquad \overline{\varphi}^{\prime a} = \overline{\varphi}^{a} + v^a.
	\end{equation}
For convenience, we define $ \overline{f}\du{A}{a} = - i T^a_{Ab} \overline{\varphi}^{\prime b} $, which contains the decay constants $ f\du{A}{a} $.
The quantum and background gauge transformations are given by 
	\begin{align} \label{eq:gauge_transforms}
	\delta_\alpha \varphi^a &= i \alpha^{A} T^{a}_{Ab} (\varphi^b + \overline{\varphi}^{\prime b}), 
	& \delta_\alpha \overline{\varphi}^{\prime a} &= 0, 
	& \delta_\alpha A_\mu^{A} &= \widetilde{D}_\mu \alpha^A,
	& \delta_\alpha \overline{A}_\mu^A &= 0, \\
	\overline{\delta}_\alpha \varphi^a &= i \alpha^{A} T^{a}_{Ab} \varphi^b, 
	& \overline{\delta}_\alpha \overline{\varphi}^{\prime a} &= i \alpha^{A} T^{a}_{Ab} \overline{\varphi}^{\prime b},
	& \overline{\delta}_\alpha A_\mu^{A} &= -f\ud{A}{BC} \alpha^B A_\mu^{C}, 
	& \overline{\delta}_\alpha \overline{A}_\mu^{A} &= \overline{D}_\mu \alpha^A, \nonumber
	\end{align}
respectively. The combined covariant derivative is denoted $ \widetilde{D}_\mu = \partial_\mu - i A_\mu - i \overline{A}_\mu $ while $ \overline{D}_\mu = \partial_\mu - i \overline{A}_\mu $ is the usual background covariant derivative. 

In theories with breaking of the gauge group, the BF $ R_\xi $ gauge-fixing condition reads
	\begin{equation} \label{eq:BF_Rxi_gauge_condition}
	G^{A}[\eta+\overline{\eta},\, \overline{\eta}] = F\ud{A}{I}[\overline{\eta}] \eta^I = \overline{D}^\mu A^A_\mu - \xi a^{AB} \overline{f}_{Ba} \varphi^a.
	\end{equation}
The gauge-fixing Lagrangian for the vectors~\eqref{eq:BFM_fixing_term} resulting from this condition is then
	\begin{equation} \label{eq:bg_fix_vector}
	\L_\mathrm{vec.}^G = -\dfrac{1}{2\xi} a^{\eminus 1}_{AB} \overline{D}^\mu \! A^A_\mu \, \overline{D}^\nu \! A^B_\nu  + (\overline{D}^\mu \!A^A_{\mu}) \overline{f}_{Aa} \varphi^a - \dfrac{\xi}{2} \varphi^a (h \overline{f}\transpose a \, \overline{f} h)_{ab} \varphi^b.
	\end{equation}
Using the quantum gauge transformation rules~\eqref{eq:gauge_transforms}, we find that the corresponding ghost Lagrangian is  
	\begin{equation} \label{eq:bg_fix_ghost}
	\L_\mathrm{gh.}^G = - \overline{\omega}_A \overline{D}^\mu ( \overline{D}_\mu \omega^A + f\ud{A}{BC} A^B_\mu \omega^C) + i \xi \,\overline{\omega}_A a^{AB} \overline{f}_{Ba} T^{a}_{Cb} (\varphi + \overline{\varphi}')^b \omega^C.
	\end{equation}
$ \L_\mathrm{gh.}^G $ contains a mass term for the ghosts proportional to $ f\, f\transpose $ responsible for separating the ghosts into massive and massless DOFs mirroring the gauge bosons. The full gauge-fixing Lagrangian of the BF $ R_\xi $-gauge is given by $ \L_\mathrm{fix}^G = \L_\mathrm{vec.}^G + \L_\mathrm{gh.}^G $. So far, so good; the real complexity arises when we go to the mass basis.

\subsection{Gauge-fixing terms in the mass basis}
The BF gauge-fixing condition~\eqref{eq:BF_Rxi_gauge_condition} is brought to the mass basis with transformation~\eqref{eq:L_transformation}. It reads 
	\begin{multline}
	L\ud{A}{B} G^{B}[\eta+ \overline{\eta},\, \overline{\eta}] = \big[ \overline{d}^\mu B^\alpha_\mu + f\ud{\alpha}{jk} \overline{V}^j_\mu V^{k\mu} - \xi \hat{a}^{\alpha \beta} \big( \chi_i f\ud{i}{\beta j} \overline{\chi}^j- i \varphi_{h\,a} t^a_{\beta b} \overline{\varphi}_h^b \big) \big] \\
	\oplus	\big[\overline{d}^\mu V^i_\mu + f\ud{i}{j\alpha} \overline{V}^j_\mu B^{\alpha\mu}+ f\ud{i}{jk} \overline{V}^j_\mu V^{k\mu} - \xi M_i \chi^i +i \xi \kappa^{ij} \varphi_a x^a_{jb} \overline{\varphi}^b \big].
	\end{multline}
The gauge-fixing Lagrangian for the vectors~\eqref{eq:bg_fix_vector} then becomes 
	\begin{align}
	\L_\mathrm{vec.}^G	&= -\dfrac{1}{2\xi} \hat{a}^{\eminus 1}_{\alpha \beta } \overline{d}^\mu B^\alpha_\mu \overline{d}^\nu B^\beta_\nu -\dfrac{1}{2\xi} \overline{d}^\mu V^i_\mu \overline{d}_\nu V^\nu_i
	+ M_i \chi_i \overline{d}^\mu V_\mu^i  - \dfrac{\xi}{2} M_i^2 \chi_i \chi^i 
	- \dfrac{1}{\xi} \overline{d}^\mu B^\alpha_\mu \hat{a}^{\eminus 1}_{\alpha\beta } f\ud{\beta}{jk} \overline{V}^j_\nu V^{k\nu} \nonumber\\
	& - \dfrac{1}{2\xi} \hat{a}^{\eminus 1}_{\alpha\beta } f\ud{\alpha}{ij} \overline{V}^i_\mu V^{j\mu} f\ud{\beta}{k\ell} \overline{V}^k_\nu V^{\ell\nu} 
	+ \big(\overline{d}^\mu B_\mu^\alpha + f\ud{\alpha}{ij} \overline{V}^i_\mu V^{j\mu} \big) \big( \chi_k f\ud{k}{\alpha \ell} \overline{\chi}^\ell- i \varphi_{h\, a} t^a_{\alpha b} \overline{\varphi}_h^b \big) \nonumber \\
	&- \dfrac{\xi}{2} \hat{a}^{\alpha \beta} \big( \chi_i f\ud{i}{\alpha j} \overline{\chi}^j- i \varphi_{h\,a} t^a_{\alpha b} \overline{\varphi}_h^b\big) \big( \chi_k f\ud{k}{\beta \ell} \overline{\chi}^\ell- i \varphi_{h\, c} t^a_{\beta d} \overline{\varphi}_h^d \big)
	-\dfrac{1}{\xi} \overline{d}^\mu V^i_\mu \overline{V}^j_\nu \big( f_{ij\alpha} B^{\alpha\nu}+ f_{ij k} V^{k\nu} \big) \nonumber\\
	& -\dfrac{1}{2\xi} \overline{V}^j_\mu \overline{V}^\ell_\nu \big( f_{ij\alpha}B^{\alpha\mu}+ f_{ijk} V^{k\mu} \big) \big( f\ud{i}{\ell\beta}B^{\beta\nu}+ f\ud{i}{\ell m} V^{m\nu} \big) + \dfrac{\xi}{2} (\varphi_a x^a_{ib} \overline{\varphi}^b) \kappa^{ij}(\varphi_c x^c_{jd} \overline{\varphi}^d) \nonumber\\
	&+ \big(M_i \chi_i -i \varphi_a x^a_{ib} \overline{\varphi}^b \big)
	\overline{V}^j_\mu \big(f\ud{i}{j\alpha} B^{\alpha\mu}+ f\ud{i}{jk}  V^{k\mu} \big) -i \overline{d}^\mu V^i_\mu \varphi_a x^a_{ib} \overline{\varphi}^b + i \xi M_i \chi^i \varphi_a x^a_{ib} \overline{\varphi}^b.
	\end{align}
Of course, all the terms involve exactly two quantum fields and must all be included when computing one-loop amplitudes. 
The scalars have only been decomposed into Higgs and GB modes where mixing terms vanish. Otherwise, the clutter would increase even more. In practice, we suspect that it is just as convenient to evaluate the remaining terms by determining the embedding of all scalars in $ \varphi $ and expanding the terms in the specific model. 

Ghost and anti-ghost fields decompose in mass states as determined by the vector bosons:
	\begin{equation}
	L\ud{A}{B} \omega^B = \binom{c^\alpha}{u^i}, \qquad
	\overline{\omega}_B (L^{\eminus1})\ud{B}{A}= \binom{\bar{c}^\alpha}{\bar{u}^i} .
	\end{equation}
The transformation of $ \overline{\omega} $ is indicative of the anti-ghost being in the dual space to the Lie algebra. It is \emph{not} the conjugate of the ghost $ \omega $. In the mass basis, the background gauge ghost term~\eqref{eq:bg_fix_ghost} reads
	\begin{align}
	\L^G_\mathrm{gh.} =\,& - \overline{c}_\alpha \overline{d}^2 c^\alpha 
	- \overline{u}_i \overline{d}^2 u^i 
	+ \overline{d}^\mu \overline{c}_\alpha f\ud{\alpha}{\beta \gamma} B^\beta_\mu c^\gamma 
	+ \overline{d}^\mu \overline{u}_i f\ud{i}{\alpha j} B^\alpha_\mu u^j
	+ \overline{d}^\mu\overline{c}_\alpha f\ud{\alpha}{ij} ( \overline{V}_\mu^i + V_\mu^i) u^j \nonumber\\
	& + \overline{d}^\mu \overline{u}_i ( \overline{V}_\mu^j + V_\mu^j) (f\ud{i}{j\alpha} c^\alpha + f\ud{i}{jk} u^k) 
	- \overline{c}_\alpha f\ud{\alpha}{ij} \overline{V}_\mu^i \overline{d}^\mu u^j 
	- \overline{u}_i \overline{V}_\mu^j (f\ud{i}{j\alpha} \overline{d}^\mu c^\alpha + f\ud{i}{jk} \overline{d
	}^\mu u^k)  \nonumber\\
	& - \overline{c}_\alpha f\ud{\alpha}{ij} f\ud{j}{\alpha k} \overline{V}_\mu^i B^{\alpha\mu} u^k 
	- \overline{u}_i \overline{V}_\mu^j B^{\beta \mu} (f\ud{i}{j\alpha} f\ud{\alpha}{\beta \gamma} c^\gamma + f\ud{i}{jk} f\ud{k}{\beta \ell} u^\ell) \nonumber\\
	&  -\overline{V}^{k\mu} ( \overline{V}_\mu^\ell + V_\mu^\ell)
	\big[\overline{c}_\alpha f\ud{\alpha}{km} f\ud{m}{\ell \beta} c^\beta  
	+\overline{c}_\alpha f\ud{\alpha}{km} f\ud{m}{\ell j} u^j 
	+\overline{u}_i f\ud{i}{km} f\ud{m}{\ell \beta} c^\beta
	\nonumber \\
	&\hspace{.2\textwidth}  +\overline{u}_i (f\ud{i}{k\alpha} f\ud{\alpha}{\ell j} + f\ud{i}{km} f\ud{m}{\ell j}) u^j \big] \nonumber\\
	&+ \xi \overline{c}_\alpha \hat{a}^{\alpha \beta} \big[\overline{\chi}_i f\ud{i}{\beta j} f\ud{j}{\gamma k} (\overline{\chi} + \chi)^k - \overline{\varphi}_{h\, a} (t_\alpha t_\beta)\ud{a}{b} (\overline{\varphi}_h + \varphi_h)^b \big] c^\gamma \nonumber \\
	&+ \xi \overline{c}_\alpha \hat{a}^{\alpha \beta} \big[M_j \chi_j f\ud{j}{\alpha i} - \overline{\varphi}_a (t_\alpha x_i)\ud{a}{b} (\overline{\varphi}+ \varphi)^b \big] u^i \nonumber \\
	& - \xi \overline{u}_i \big[f\ud{i}{\alpha j}M_j (\overline{\chi} + \chi)^j + \overline{\varphi}_a (x^i t_\alpha)\ud{a}{b} (\overline{\varphi} + \varphi)^b \big] c^\alpha \nonumber \\
	&- \xi \overline{u}_i \big[\delta\ud{i}{j} M_j^2 - i f\ud{i}{a} x^a_{jb} (\overline{\varphi} + \varphi)^b - i f_{ja} \kappa^{ik} x^a_{kb} \overline{\varphi}^b + \overline{\varphi}_a (x^i x_j)\ud{a}{b} (\overline{\varphi} + \varphi)^b \big] u^j  
	\end{align}
Some of these terms involve three quantum fields (including ghosts) and can safely be ignored at one-loop order. 
Observe that $ \mathbf{c} $ remains massless as expected, while on the other hand, $ \mathbf{u} $ gets a mass term, which in the Feynman gauge ($ \xi= 1 $) coincides with that of the massive vector fields, $ V_\mu $.  
Between the ghost and vector terms of the BF gauge fixing terms, the reader will no doubt agree with our assessment that the partially fixed BF gauge leads to significantly simpler gauge-fixing terms: imagine what the analog to the gauge-fixing terms~(\ref{eq:toy_model_vector}, \ref{eq:toy_model_ghost}) would look like in the ordinary BF gauge.

\sectionlike{References}
\vspace{-10pt}
\bibliography{References} 
\end{document}